\documentclass[sigconf]{acmart}

\usepackage{hyperref}
\usepackage{acmart-taps}
\usepackage{tabularx}
\usepackage{graphicx}
\usepackage{array}
\usepackage{booktabs}  
\usepackage{color}     
\usepackage{multirow}
\usepackage{makecell}
\usepackage{colortbl} 
\usepackage{enumitem}
\usepackage{subcaption}
\usepackage{dirtytalk}
\usepackage{soul} 
\usepackage{xcolor} 

\usepackage{stfloats} 

\newcommand{\rv}[1]{\textcolor{black}{#1}}

\usepackage{listings}

\lstdefinestyle{jsonstyle}{
    keywordstyle=\color{blue},
    stringstyle=\color{red},
    numberstyle=\color{purple},
    commentstyle=\color{gray},
    breaklines=true,
    breakatwhitespace=false,
    showstringspaces=false,
    morestring=[b]",
    morecomment=[s]{/*}{*/},
    morecomment=[l]//,
    morekeywords={true,false,null},
    literate=
        *{:}{{{\color{blue}:}}}{1}
        {,}{{{\color{blue},}}}{1},
    aboveskip=5pt,
    belowskip=5pt,
    xleftmargin=0pt,
    xrightmargin=0pt,
}

\lstdefinelanguage{json}{
   morestring=[b]",
   morestring=[d]'
}

\lstdefinestyle{customjson}{
   backgroundcolor=\color{white}, 
   basicstyle=\ttfamily\footnotesize, 
   breakatwhitespace=false,         
   breaklines=true,                 
   captionpos=b,                    
   keepspaces=true,                 
   numbers=left,                    
   numbersep=5pt,                  
   showspaces=false,                
   showstringspaces=false,
   showtabs=false,                  
   tabsize=2,
   language=json,
   stringstyle=\color{blue},
   keywordstyle=\color{purple}
}

\definecolor{revisioncolor}{rgb}{0.0, 0.0, 1.0}

\AtBeginDocument{%
  }

\setcopyright{acmcopyright}
\copyrightyear{2024}
\acmYear{2024}
\setcopyright{rightsretained}
\acmConference[UIST '24]{The 37th Annual ACM Symposium on User Interface Software and Technology}{October 13--16, 2024}{Pittsburgh, PA, USA}
\acmBooktitle{The 37th Annual ACM Symposium on User Interface Software and Technology (UIST '24), October 13--16, 2024, Pittsburgh, PA, USA}
\acmDOI{10.1145/3654777.3676359}
\acmISBN{979-8-4007-0628-8/24/10}

\usepackage{soul}     
\usepackage{color}    
\usepackage{xspace}     
\usepackage{xpunctuate} 
\usepackage{natbib}
\usepackage{transparent}
\usepackage{tikz}
\usepackage{setspace}

\newcommand{\sys}{\textit{Aletheia}}

\newcommand{\ie}{{i.e.,}\xspace}
\newcommand{\eg}{{e.g.,}\xspace}

\definecolor{lightpink}{RGB}{237,157,202}
\definecolor{lightred}{RGB}{210,121,121}
\definecolor{lightorange}{RGB}{230,170,50}
\definecolor{lightgold}{RGB}{210,194,121}
\definecolor{lightgreen}{RGB}{121,210,121}
\definecolor{lightaqua}{RGB}{121,206,210}
\definecolor{lightblue}{RGB}{121,124,210}
\definecolor{lightpurple}{RGB}{153,102,255}
\definecolor{red}{RGB}{178,34,34}
\definecolor{gray}{RGB}{166,166,166}
\definecolor{peach}{RGB}{255,218,185}

\definecolor{Amber}{RGB}{255,190,11}
\definecolor{Pantone}{RGB}{251,86,7}
\definecolor{Rose}{RGB}{255,0,110}
\definecolor{Violet}{RGB}{131,56,236}
\definecolor{Azure}{RGB}{58,134,255}
\definecolor{mypurple}{RGB}{131,56,236}
\definecolor{myred}{RGB}{251,86,7}
\definecolor{mydarkred}{RGB}{252,102,32}
\definecolor{mydarkpurple}{RGB}{144,77,238}

\newcommand{\measureColor}{Violet}
\newcommand{\valueColor}{Azure}
\newcommand{\aggregationColor}{lightaqua}
\newcommand{\focusColor}{lightgreen}
\newcommand{\subspaceColor}{lightorange}
\newcommand{\identifiedColor}{Pantone}
\newcommand{\claimExampleSpacing}{\vspace{4px}}

\newcommand{\claimSpecColorNEW}[3][1=]{\setulcolor{#3}{\ul{#1}}}
\newcommand{\claimVarStyle}[1]{\texttt{\{#1\}}}
\newcommand{\claimVarStyleNEW}[1]{\textnormal{#1}}

\begin{document}

\title{``The Data Says Otherwise'' -- Towards Automated Fact-checking and Communication of Data Claims}

\author{Yu Fu}
\email{fuyu@gatech.edu}
\affiliation{%
  \institution{Georgia Tech}
  \city{Atlanta}
  \country{United States}}

\author{Shunan Guo}
\email{sguo@adobe.com}
\affiliation{%
  \institution{Adobe Research}
  \city{San Jose}
  \country{United States}}

\author{Jane Hoffswell}
\email{jhoffs@adobe.com}
\affiliation{%
  \institution{Adobe Research}
  \city{San Jose}
  \country{United States}}

\author{Victor S.Bursztyn}
\email{soaresbu@adobe.com}
\affiliation{%
  \institution{Adobe Research}
  \city{San Jose}
  \country{United States}}

\author{Ryan Rossi}
\email{ryrossi@adobe.com}
\affiliation{%
  \institution{Adobe Research}
  \city{San Jose}
  \country{United States}}

    \author{John Stasko}
\email{john.stasko@cc.gatech.edu}
\affiliation{%
  \institution{Georgia Tech}
  \city{Atlanta, GA}
  \country{United States}}



\begin{abstract}
Fact-checking data claims requires data evidence retrieval and analysis, which can become tedious and intractable when done manually. This work presents \sys{}, an automated fact-checking prototype designed to facilitate data claims \emph{verification} and enhance data evidence \emph{communication}. For verification, we utilize a pre-trained LLM to parse the semantics for evidence retrieval. To effectively communicate the data evidence, we design representations in two forms: data tables and visualizations, tailored to various data fact types. Additionally, we design interactions that showcase a real-world application of these techniques.  We evaluate the performance of two core NLP tasks with a curated dataset comprising 400 data claims and compare the two representation forms regarding viewers' assessment time, confidence, and preference via a user study with 20 participants. The evaluation offers insights into the feasibility and bottlenecks of using LLMs for data fact-checking tasks, potential advantages and disadvantages of using visualizations over data tables, and design recommendations for presenting data evidence.
\end{abstract}



\maketitle

\section{Introduction}

Imagine skimming through an ESPN analysis comparing two athletes' performances, reading a New York Times article about COVID trends, or browsing Fox News coverage on the upcoming U.S. election. These articles likely contain many data claims.
However, data claims can contain inaccuracies from various sources. Errors and omissions might arise from oversights during the composition phase, where analysts engage in data analysis and manually transcribe insights into textual form~\cite{Chen_2022_CrossData}. Additionally, the frequent updates in data can exacerbate discrepancies between the data and the claims made. More alarmingly, bad actors may deliberately manipulate or fabricate data facts to advance specific agendas or propaganda. Regardless of intent, such flawed data claims contribute to a flood of misinformation that inundates and contaminates our information ecosystem, potentially misleading the public.

A standard practice to mitigate misinformation is through fact-checking, a process of assessing the veracity of textual claims based on authoritative or trusted evidence. Typically undertaken by professional fact-checkers within news organizations, 
fact-checking has long held a vital role in upholding the accuracy and integrity of information~\cite{graves2016deciding}. However, with manual fact-checking challenged by the increasing column of information production and dissemination, both practitioners and researchers are turning to advanced technology, notably \emph{automated fact-checking (AFC)}~\cite{Guo_2022_survey_automated_factchecking, Adair2017ProgressT}.

Automated fact-checking applies to many scenarios, especially in news platforms and social media. Researchers in computational journalism~\cite{Cohen_2011_callforarmstodatabase, flew_spurgeon_daniel_swift_2012} have advocated for sophisticated technologies to enhance and facilitate fact-checking tasks. Such tools can serve multiple roles for fact-checkers and journalists alike, and empower news readers to critically audit the content they consume~\cite{Fu_2023_Morethan}.
Recently, the data mining and natural language processing (NLP) communities have contributed to a growing body of research to address this demand~\cite{Wu_2014_iCheck, Wu_2017_ComputationalFactChecking}, with a particular emphasis on downstream tasks~\cite{zubiaga2018detection,kuccuk2020stance, kotonya2020explainable}, domain-specific requirements~\cite{kotonya-toni-2020-explainable-automated, wadden2020fact}, and the provision of annotated claim-evidence datasets~\cite{thorne-etal-2018-fever, hanselowski2019richly} for model training. 
The majority of prior work has mainly concentrated on text-based evidence, wherein claims are verified by cross-referencing them with a textual corpus of established facts, including sources like Wikipedia pages~\cite{schuster2021get} and scientific articles~\cite{wadden2022multivers}.

Data claims, also known as numeric or statistical claims~\cite{Cao_2019_ExtractingStatisticalMentions}, use natural language to describe facts/insights derived from structured data and/or statistics. We posit that the veracity of data claims is intrinsically tied to specific quantitative datasets, leading to a divergence in tasks from the conventional text-based fact-checking pipeline.
This divergence can manifest at various stages, primarily evidence retrieval and presentation.
While some existing solutions for automated fact-checking translate natural language into SQL queries~\cite{jo2019verifying,Karagiannis_2020_Scrutinizer}, these methods often struggle to address more complex insights (\eg anomalies, trends, associations, etc.).
These methods also require a pre-established knowledge base or set of SQL query candidates tailored to the dataset, limiting their applicability to new datasets.
Furthermore, despite recognizing the importance of providing corroborating justification~\cite{Guo_2022_survey_automated_factchecking, Vallayil_2023_Explainability}, research on data evidence presentation and optimal representation forms for diverse data claims is scarce. Effective presentation of data evidence can not only bolster persuasiveness but also empower viewers to pinpoint inconsistencies between evidence and outcomes.

Thus, this work tackles two primary research questions: \textbf{Q1}.~\emph{``how do we enable out-of-the-box automated fact-checking for data claims?''} and \textbf{Q2}.~\emph{``how do we effectively represent and communicate data evidence?''} 
%
Building upon an established fact-checking framework~\cite{Guo_2022_survey_automated_factchecking}, we first propose a modified automated data fact-checking framework (\autoref{fig:framework}) comprising six components: \emph{data claim detection}~(C1), 
\emph{text-to-data mapping}~(C2),
\emph{data evidence retrieval}~(C3), 
\emph{verdict determination \& presentation}~(C4),
\emph{data evidence presentation}~(C5),
and \emph{end user interaction}~(C6). 
We design and develop a prototype fact-checking system, \sys, to integrate these components, serving as both a proof-of-concept for our proposed framework and a design prototype that showcases its feasibility for practical applications.

This pipeline uses GPT models for downstream NLP tasks, leveraging its innate semantic parsing abilities~\cite{zhuo2023robustness, schucher2022power}. Informed by a content analysis of data claims in sixteen real-world articles, our prompting pipeline uses seven steps~(\autoref{fig:prompt_pipeline}) to transform claims into data fact specifications~\cite{ding2019quickinsights, Wang_2020DataShot, Shi_2021_Colliope}.
This transformation enhances the transparency and interpretability of the connections between claims, the pertinent data subsets, and the derived insights. 

For more effective data evidence communication, we introduce twenty-six data evidence representations across both data tables and visualizations for thirteen data fact types. To improve \sys's practical utility, we incorporate interactions to facilitate stakeholders' fact-checking needs, such as overriding AI-induced mistakes. 

We evaluate \sys{}'s two key components: the performance of the core steps in our \emph{claim-to-data} transformation pipeline (\textbf{Q1}), and the effectiveness of the data evidence presentation when comparing the data table and visualization (\textbf{Q2}). In particular, we assess the backend pipeline on a manually curated dataset of 400 claims across various types (\autoref{section5}), demonstrating the LLM's promising capabilities in classifying data facts and converting natural language data claims to data fact specifications. 
Regarding \textbf{Q2}, we conducted a mixed-method user study with 20 participants tasked with reviewing data claims (\autoref{userstudy}). Our findings indicate that visualization charts outperform data tables in terms of the reviewing time for most data fact types (7 out of 13), enhance participant confidence across all data fact types, and are preferred in the majority. Drawing from our findings, we ultimately put forth four general design recommendations for effectively presenting data evidence.

\section{Related Work}\label{section:rw}
This work is driven by prior research in automated fact-checking, existing strategies on justification presentation, and techniques for visually linking between text, tables, visualizations, and data.
\subsection{Automated Fact-checking}
Automated fact-checking has garnered significant attention in the NLP community to help counter misinformation and disinformation. Extensive research effort has been dedicated towards downstream tasks, including claim detection~\cite{hassan2017toward, hassan2015detecting}, evidence retrieval~\cite{kuccuk2020stance, hardalov2022survey}, verdict prediction~\cite{tan2023multi2claim, popat2018declare}, justification production~\cite{nguyen2018believe, jacovi2020towards}, and more. \mbox{\citeauthor{Guo_2022_survey_automated_factchecking}~\cite{Guo_2022_survey_automated_factchecking}} consolidate these tasks into a cohesive framework outlining the essential components for automated fact-checking systems. \textcolor{black}{We refined this framework specifically for data claims, leading to the creation of \sys{}}.  

\rv{Traditionally}, automated fact-checking systems was grounded on knowledge bases, \rv{verifying claims} against a textual corpus of accumulated facts (e.g., Wikipedia pages~\cite{schuster2021get}, scientific articles~\cite{wadden2022multivers}, or knowledge graphs~\cite{tchechmedjiev2019claimskg}). These systems rely on pre-established, reliable information sources to identify related supporting claims as evidence and determine the veracity based on the coherence with the evidence. In contrast, our work focuses on data claims that are not explicitly in the knowledge base but can be inferred from structured data tables. Data tables are a ubiquitous medium for storing information across various applications, and practitioners(e.g., data analysts, business analysts, etc.) often create text reports to summarize insightful statistics. 

Previous research in text-to-data matching has tackled similar challenges, linking entities in text paragraphs to data tables through semantic parsing~\cite{thorne2017extensible, li2020deep, mudgal2018deep, herzig2020tapas, yin2020tabert}. \rv{Additionally, chart reasoning techniques have been applied to fact-checking applications, focusing on verifying the correctness of data statements with a given chart image. For instance, Akhtar et al. introduced two baseline datasets~\cite{akhtar2023chartcheck, akhtar2023reading} for generating explainable fact-checking results over chart images. Most of these methods operate within supervised settings, which require expensive training on extensive documents, data tables, and chart images.} Alternatively, unsupervised solutions~\cite{ahmadi2022unsupervised} often suffer from limited scalability and unstable performance~\cite{wang2022promptem}. Another line of research addresses this problem by translating natural language claims into SQL queries and validating the claimed values against the queried results. For example, AggChecker~\cite{jo2019verifying} maps data claims to a probability distribution over a set of candidate SQL queries. While this method operates in an unsupervised manner, expanding the system to accommodate new datasets requires updates to the query candidates and probabilistic models to account for new table schemas. Similarly, Scrutinizer~\cite{Karagiannis_2020_Scrutinizer} employs an NL-to-SQL translation strategy but integrates an additional mixed-initiative pipeline that permits input from human experts to guide the translation process. However, the initial translation model relies on machine learning classifiers trained exclusively on the schema of the input data table, thereby constraining its adaptability to new datasets. 

Considering the broad accessibility of pretrained LLMs (e.g., GPTs) and the proven ability in initial fact-checking trials~\cite{hoes2023using, stammbach2020fever, buchholz2023assessing}, this study delves into a fact-checking solution harnessing the integrated capability of pretrained LLM. Our primary goal is to explore an ``out-of-the-box'' fact-checking solution designed for non-expert practitioners, such as journalists and business analysts, who often lack the resources for model training and may not possess an in-depth understanding of complex fact-checking models.

\subsection{Verdict \textcolor{black}{and Justification} Presentation}
When assessing verdicts through automated approaches, it is crucial to communicate its fact-checking decisions to the fact reviewers with comprehensible justifications~\cite{graves2018understanding, kotonya2020explainable, Guo_2022_survey_automated_factchecking, Vallayil_2023_Explainability, Das_2023_stateofHuman_ceteredNLPFact_checking}. Consequently, effectively presenting fact-checking results emerges as a vital research aspect that culminates at the end of the fact-checking process~\cite{spina2023human}. The primary method of conveying verdicts involves the use of veracity indicators~\cite{amazeen2018correcting}, \ie graphical elements succinctly encapsulating the veracity of claims on truth scales. For instance, color codings are commonly applied to present fact-checking outcomes in a variety of fact-checking research endeavors~\cite{rezgui2021automatic, jo2019verifying, thorne-etal-2018-fever}. Public-facing fact-checking platforms, such as PolitiFact and Snopes, often include more comprehensive fact-check ratings that encompass not only varying levels of claim truthfulness but also categories such as scam, outdated, or research-in-progress to enhance credibility. \rv{ClaimViz~\cite{rony2020claimviz} presents a visual analytics system that supports journalists in reviewing large amounts of factual claims and identifying check-worthy ones.}In an effort to provide guidance on effective presentation of verified information to fact-checking report readers, \mbox{\citeauthor{hettiachchi2023designing}~\cite{hettiachchi2023designing}} identified six critical design elements in fact-checking reports and studied their impact on improving the credibility and presentation of the reports with crowd-sourced experiments. 

\textcolor{black}{In the pursuit of enhancing the explainability of automated fact-checking systems, automated fact-checking research has employed different  approaches~\cite{Guo_2022_survey_automated_factchecking, Das_2023_stateofHuman_ceteredNLPFact_checking, Vallayil_2023_Explainability}, including summarization (extractive and abstractive)~\cite{kotonya2020explainable, atanasova-etal-2020-generating-fact}, logic-based~\cite{Chen_Bao_Sun_Zhang_Chen_Zhou_Xiao_Li_2022}, attention-based~\cite{popat-etal-2018-declare, shu_cui_wang_lee_liu_2019}, and counterfactual~\cite{xu-etal-2023-counterfactual} methods. \mbox{\citeauthor{Vallayil_2023_Explainability}~\cite{Vallayil_2023_Explainability}} specifically examines the application of explainable AI (XAI) to automated fact-checking, highlighting significant challenges existing in multiple aspects, including the current lack of datasets that facilitate the explanations production and the ambiguity surrounding different concepts and taxonomy (e.g., global vs. local explainability). More recent studies (e.g.\cite{Althabiti_2023_AFC_XAIDATA}) aim to provide datasets for explainable fact-checking. However, it is worth noting existing explainable fact-checking research also predominately resolves around} claims and evidence presented in unstructured text, whereas our work centers on data-driven claims and evidence rooted in structured quantitative information that requires distinct forms of presentation.

\subsection{Linking Data to Visual Representations}
When reviewing data-driven claims, identifying relevant data sources serves as the cornerstone for assessing the veracity~\cite{nguyen2016state, juneja2022human}. Consequently, effectively communicating the underlying data to viewers during the process is pivotal in developing data fact-checking systems. The HCI community has made substantial contributions in advancing the realm of efficient data communication and content consumption within data documents using other visual representations, including data tables and visualization charts. For instance, \mbox{\citeauthor{kong2014extracting}~\cite{kong2014extracting}} developed an interactive document viewer with the reference among text, tables, and visualization charts established by crowdsourced workers. \mbox{\citeauthor{kim2018facilitating}~\cite{kim2018facilitating}} automated the association between text and table cells using NLP techniques, enabling the interactive highlighting of relevant table cells in response to user-selected sentences. \mbox{\citeauthor{badam2018elastic}~\cite{badam2018elastic}} proposed to connect text and tables through the generated contextual visualizations to enhance the reading experience. \citeauthor{latif2021kori} introduced Kori~\cite{latif2021kori}, a mixed-initiative interface designed to facilitate the authoring process of interactive data documents by offering both recommendations for linking text with charts and manual construction of references. These techniques effectively link textual content with predefined tables or visualizations embedded in the same document. In our fact-checking context, we consider the entire dataset behind the scenes, with the audience exposed solely to the textual content.

Within this context, \citeauthor{Chen_2022_CrossData} developed CrossData~\cite{Chen_2022_CrossData}, an authoring assistance system that retrieves backend data and presents it in table or visualization form during the document authoring process. CrossData primarily focuses on providing rich interactions to facilitate author-driven associations between text, tables, and visualizations during document creation. Conversely, our work is centered on devising means to effectively communicate data evidence relevant to claims. \mbox{\citeauthor{zhi2019linking}~\cite{zhi2019linking}} highlighted the positive impact of linking visualizations and text in storytelling on aspects such as comprehension, engagement, and recall. However, the objectives differ regarding the communication of data evidence, encompassing efficiency, data consistency, and user confidence in the verdict. In our work, we explore two visual representations --- data tables and visualization charts, as means to effectively present data evidence. We assess their impact on efficiency, user confidence, and preference during the claim review tasks through user studies.

\section{A Framework for Data Claim Fact-Checking and Communication}\label{sec:scope}
\begin{figure*}[t]
    \centering
    \includegraphics[width=0.95\textwidth]{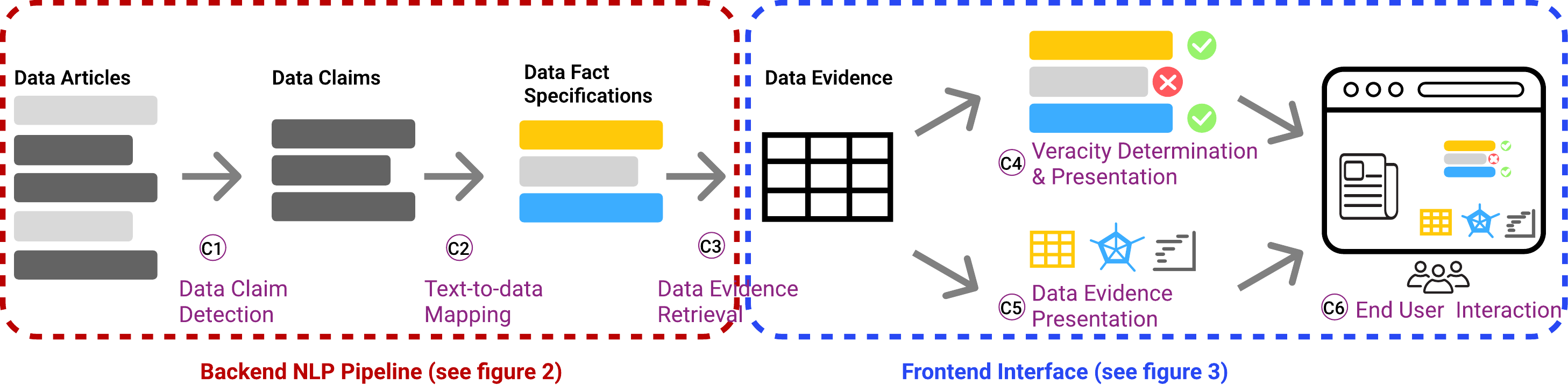}
    \caption[]{An overview of our modified framework for automated data claim fact-checking and communication, based on \citeauthor{Guo_2022_survey_automated_factchecking}'s framework~\cite{Guo_2022_survey_automated_factchecking}. The process begins by extracting data claims from data articles. These claims are mapped into data fact specifications designed to fetch pertinent evidence. This evidence not only aids in determining the veracity of the associated data claim but also serves as the justification for the verdict. The initial three components form a pipeline for the NLP tasks. This NLP pipeline (\autoref{fig:prompt_pipeline}) underpins \sys's backend. The last three components connect to \sys's interface (\autoref{fig:Aletheia}).}
    \Description{Illustrative diagram of a modified framework for automated data claim fact-checking and communication, building upon the model proposed by Guo in 2022. The depicted framework initiates with the extraction of data claims from articles, followed by the mapping of these claims into data fact specifications to gather relevant evidence. This evidence is critical for verifying the claims and providing a foundation for the final judgment. The diagram highlights a three-component pipeline focusing on Natural Language Processing (NLP) tasks that support the backend of a system referred to as 'the system.' The subsequent three components are shown linking to the system's user interface, indicating the integration of backend processing with frontend interaction.}
    \label{fig:framework}
\end{figure*}

Although the four-part NLP framework formulated by \citeauthor{Guo_2022_survey_automated_factchecking}~\cite{Guo_2022_survey_automated_factchecking} encapsulates the essence of fact-checking tasks, the downstream tasks are rooted in knowledge-based fact-checking research~\cite{wadden2022multivers,schuster2021get, tchechmedjiev2019claimskg} focusing on claims sourced from qualitative evidence. Our work, in contrast, focuses on fact-checking \textbf{data claims}: natural language sentences with one or more facts from \emph{quantitative} information.

Fact/knowledge-based claims and data claims differ in the nature of their evidence. The former, like the statement ``\textit{The director of the movie `Oppenheimer' also directed `Interstellar',}'' are typically verifiable through historical records, direct evidence, or established knowledge. In contrast, a data claim, such as ``\textit{The total gross of `Oppenheimer' accounts for 20\% of the worldwide box office gross of all films directed by Christopher Nolan,}'' relies on retrieving and aggregating a collection of data points across specific measurements (e.g., gross), aligning with the claimed insight type (i.e., proportion) and matching the asserted value against the gathered data.

While sharing parallel goals with conventional automated fact-checking endeavors, our focus on data claims necessitates a distinct approach to processing claims and conveying data evidence. To align closely with our focus, we have adapted the NLP framework proposed by \citeauthor{Guo_2022_survey_automated_factchecking}~\cite{Guo_2022_survey_automated_factchecking} to model the automated fact-checking and communication process for data claims. This modified framework (\autoref{fig:framework}), comprises the following six components: 

\subsubsection*{\textbf{C1. Data Claim Detection}} 
Our modified framework starts with the extraction of individual data claims from data articles or reports. \textcolor{black}{This focus diverges from the conventional NLP approach to claim identification, which primarily concentrates on assessing the `check-worthiness'  and `checkability' of claims~\cite{Guo_2022_survey_automated_factchecking, nakov2021automated, Das_2023_stateofHuman_ceteredNLPFact_checking, Atanasova2018OverviewOT}. }

\subsubsection*{\textbf{C2. Text-to-data Mapping}}
The extracted data claims are transformed into corresponding data fact specifications, with \textit{Text-to-data Mapping} accommodating the distinct evidence retrieval process and data aggregations involved in data claims. Researchers have developed frameworks and taxonomies to encapsulate the diversity and characteristics of such `data facts/insights'~\cite{Amar_2005_low_level_analytic_activity, Yi_2008_insights, North_2006_vis_insight, Law_2020_AutoInsights, Chen_2009_effectiveInsight}. 
We use the term \textbf{\textit{data fact}} as the granular representation of the data insight extracted from the textual claim. Particularly, we adopt the specifications of data fact from previous research~\cite{Shi_2021_Colliope, Wang_2020DataShot} to define the core data and insight within the claim descriptors such as data fact types, subspace, value, aggregation, measure, etc.

\subsubsection*{\textbf{C3. Data Evidence Retrieval}}
The data fact specifications are used to retrieve pertinent \emph{\textbf{data evidence}}, \ie the subset of data directly related to the claim. This process differs significantly from conventional NLP-focused evidence retrieval that searches for credible information from large text corpus/knowledge bases~\cite{schuster2021get,wadden2022multivers, tchechmedjiev2019claimskg} or incorporating additional metadata~\cite{wang_2017_liar}. In our work, data evidence consists of structured formats of numeric information. For instance, fact-checking a data claim about COVID-19 trends can utilize datasets from official/authoritative sources (e.g., WHO~\cite{COVID_data}). Thus, our work assumes the availability of a specific dataset to check against and concentrates on retrieving the relevant data evidence.

\subsubsection*{\textbf{C4. Veracity Determination \& Presentation}}
Unlike knowledge-based fact-checking approaches that rely on pre-existing credible text excerpts, our method employs the procured data evidence and the associated data operations. We assess the veracity of a data claim by comparing the computed values/statistics with those claimed in the text.
Note that veracity assessment can depend on three fundamental dimensions: \emph{clarity} at the linguistic level, \emph{consistency} with the data at the factual level, and \emph{conformity} with the logic at the reasoning level. While our work encompasses steps to disambiguate the linguistic expressions of data claims, our scope remains on the factual level --- ensuring that the textual description aligns with the actual data but not considering veracity at the reasoning level.

\subsubsection*{\textbf{C5. Data Evidence Presentation}}
The data evidence and operations need to be communicated to users for verdict justification. Unlike presentations used for qualitative evidence~\cite{Guo_2022_survey_automated_factchecking, Das_2023_stateofHuman_ceteredNLPFact_checking, Vallayil_2023_Explainability}, our work involves quantitative evidence, such as the subset of data and the statistical logic/rules associated with claimed data insights.
This departure leads us into the realm of HCI and data visualization, which has received limited attention in previous studies~\cite {hettiachchi2023designing}. In this work, we aim to investigate innovative methods for more effective communication of data evidence (introduced in \autoref{visdesign}).

\subsubsection*{\textbf{C6. End User Interaction}}
Previous fact-checking frameworks have often overlooked human involvement, which can improve fact-checking outcomes by clarifying semantics, correcting AI errors, and making the fact-checking outcomes more actionable. 
With diverse end-users, their interaction needs can vary greatly; for instance, authors or editors may revise a data article to ensure accuracy, whereas fact-checkers aim to explain problematic data claims to a broader audience. Our work addresses these needs by proposing user interactions with AI-driven fact-checking tools that can operationalize preceding components effectively.





\section{Aletheia}


We have developed a prototype, \sys{}, to encapsulate these six components in our framework. \sys{} is an interactive system with an LLM-based backend and a web-based frontend interface. 



\begin{figure*}[ht]
    \centering
     \footnotesize
    \includegraphics[width=0.90\textwidth]{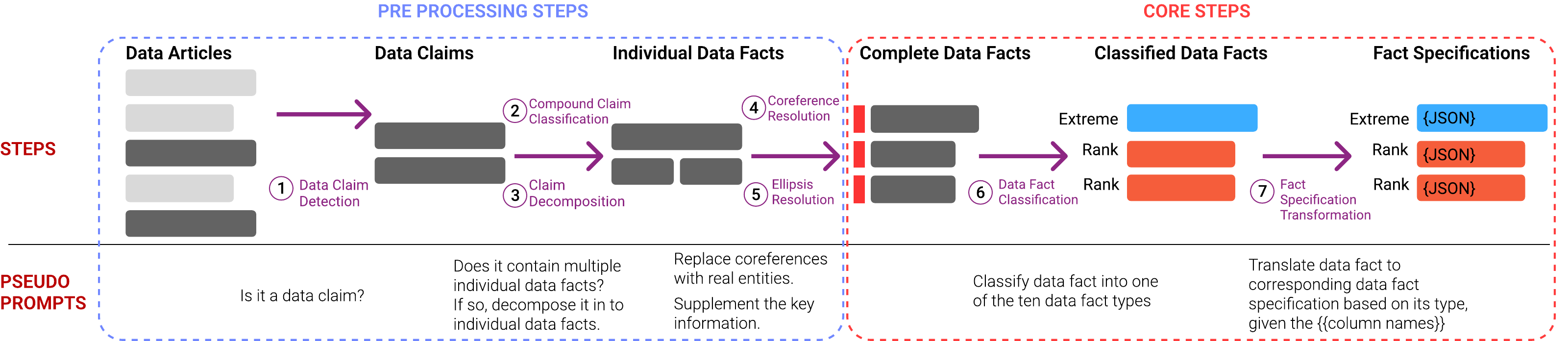}
    \caption{Overview of our LLM-based pipeline, which takes in a data article and outputs JSON specifications used to retrieve data evidence for each data claim. Seven steps are chained. The first five steps form the ``pre-processing phase'', which transforms the input text first into individual data claims~(S1) with compound claim identification~(S2), then into distinct data facts~(S3) with coreference resolution~(S4) and ellipsis resolution~(S5). These data facts are further processed in the last two steps, the ``core steps'' of our prompt pipeline, which classify the types of data facts~(S6) before converting them into data fact specifications used for retrieving the pertinent data evidence~(S7).
    }
    \Description{Flowchart diagram illustrating a Large Language Model (LLM)-based pipeline designed for processing data articles and outputting JSON specifications to support data claim verification. The pipeline is divided into seven sequential steps, with the initial five steps comprising the pre-processing phase. This phase includes: 1) Extraction of individual data claims, 2) Identification of compound claims, 3) Segregation into distinct data facts, 4) Coreference resolution, and 5) Ellipsis resolution. These processed data facts then proceed to the core steps of the pipeline: 6) Classification of data fact types, and 7) Conversion into data fact specifications for evidence retrieval. The diagram emphasizes the structured approach from initial claim extraction to the final evidence-gathering stage.}
    \label{fig:prompt_pipeline}
\end{figure*}
\subsection{Backend LLM-based Pipeline}\label{sec:prompt-pipeline}

A pivotal component in \sys's framework is extracting data claims and retrieving the pertinent data evidence (C1--C3). This process converts a data-rich article into a series of data fact specifications that can be readily employed for data retrieval.


\subsubsection{Design Implications for Prompt Pipeline}
We conducted a qualitative content analysis of real-world data claims to guide us in defining subtasks for our prompt design. We examined sixteen real-world data articles from various topics and sources, manually extracting 108 data claims for thematic analysis. This step involved identifying key themes, refining initial codes, and engaging in iterative discussions to agree on the final codes and themes. The data claim corpus and the established codes are accessible in the supplemental materials. Our analysis yielded four main findings, each leads to an implication for our prompt pipeline design:

\begin{table}[ht]
\centering
\footnotesize
\renewcommand{\arraystretch}{1.5}
\begin{tabular}{@{}p{0.43\columnwidth}p{0.51\columnwidth}@{}}
\toprule
    \textbf{Findings} & \textbf{Implications} \\ \midrule
    F1. Data claims often include more than one data fact. & I1. Decompose compound data claims into multiple independent, individual data facts. \\
    F2. Inconsistent/ambiguous terms may be used to describe attributes. & I2. Provide the reference dataset's attribute list to improve identification accuracy. \\
    F3. The contextual information determining the subspace may be omitted. & I3. Infer and supplement key information based on context to determine subspace. \\
    F4. Coreferences are generally used to represent real-world entities. & I4. Replace coreferences with actual entities based on context. \\ \bottomrule
\end{tabular}
\caption{Findings and implications for prompt design}
\label{tab:findings_implications}
\vspace{-8pt}
\end{table}

\subsubsection*{\textbf{Summary:}} 
The complexity and ambiguity inherent in real-world data claims require that they be decomposed into simpler, more verifiable facts, and disambiguated through more precise referencing and scoping. For example, to verify the claim, ``\textit{Prices tumbled 1.1\% year-on-year, logging their first annual decline since June 2020}'' from a news article~\cite{ward_glenton_2023}, the process involves \textbf{\emph{I1}} to dissect the claim into two distinct data facts (i.e., \textit{`tumbled 1.1\%'} and \textit{`first annual decline'}); \textbf{\emph{I2}} and \textbf{\emph{I3}} to specify what `prices' refer to, in this case, `housing prices,' and to define their geographical and temporal scope; and \textbf{\emph{I4}} to ensure that `their' refers to `housing prices.'

\subsubsection{Developing an LLM-based Pipeline}
Our content analysis indicates that fact-checking data claims require several NLP tasks. Training models from scratch for each task requires immense amounts of data, computational resources, and time. Consequently, we employed a pre-trained LLM (i.e., GPT-3.5) to address this multifaceted NLP challenge because it permits flexible pipeline assembly through iterative prompt engineering~\cite{White_2023_PromptPatternCatelogtoEnhancePromptEngineering, Liu_2023_pretrainePromt, Brown_2020_fewshots}. Such flexibility aligns well with our primary objective: not necessarily to attain peak performance but to investigate the viability of a fully–automated fact-checking pipeline, formulate logical, coherent steps, and garner insights that can inform subsequent end-to-end optimization.

\subsubsection*{\textbf{Pipeline Components}}
Our proposed NLP pipeline consists of seven chained steps. Due to the uncertainty in both our pipeline structure and the LLM's ability to handle each task during our initial trial-and-error phase~\cite{dang2022prompt}, we adopted the notion of LLM-chaining proposed by \citeauthor{Wu_CHI22_AIChains}~\cite{Wu_CHI22_AIChains}. Rather than burdening the LLM with the end task of generating data fact specifications directly from data articles, we split this complex task into smaller, more tractable sub-tasks. This approach enabled us to build the pipeline with greater control and transparency of its intermediate steps. We also use NLTK~\cite{NLTK} for sentence tokenization pre-processing. \autoref{fig:prompt_pipeline} illustrates the pipeline and briefly explains each step within it. 

\begin{table*}[ht]
    \footnotesize
    \centering
    \renewcommand{\arraystretch}{0.9}
\begin{imageonly}
    \begin{tabular}{>{\arraybackslash}m{1.25cm} m{4.64cm} m{8.34cm}}
        \centering \textbf{Fact Types} &
        \centering \textbf{Claim Examples} &
        \centering \textbf{JSON specifications} 
        \tabularnewline
        \toprule 
        Value
        & 
        \claimExampleSpacing
        The
        \claimSpecColorNEW[average]{\claimVarStyle{Agg}}{\aggregationColor}
        \claimSpecColorNEW[IMDB score]{\claimVarStyle{M}}{\measureColor}
        for
        \claimSpecColorNEW[horror]{\claimVarStyle{$S_1$}}{\subspaceColor}
        \claimSpecColorNEW[movies]{\claimVarStyle{ID\_key}}{\identifiedColor}
        released in
        \claimSpecColorNEW[2020]{\claimVarStyle{$S_2$}}{\subspaceColor}
        is
        \claimSpecColorNEW[6.7]{\claimVarStyle{V}}{\valueColor}
        \hspace{-2px}.
        \claimExampleSpacing
        &    
        \texttt{\{%
            "measure": \textcolor{\measureColor}{\claimVarStyleNEW{``IMDB score''}},
            "value":~\textcolor{\valueColor}{\claimVarStyleNEW{6.7}},
            "aggregation":~\textcolor{\aggregationColor}{\claimVarStyleNEW{``average''}},
            "subspace":~[\textcolor{\subspaceColor}{\claimVarStyleNEW{\{"genre"="horror"\}}},\textcolor{\subspaceColor}{\claimVarStyleNEW{\{"year"=2020\}}}],
            "identifier\_key":~\textcolor{\identifiedColor}{\claimVarStyleNEW{``movies''}}%
        \}}
        \\
        \midrule
        Proportion 
        &
        \claimExampleSpacing
        In \claimSpecColorNEW[2013]{\claimVarStyle{$S_1$}}{\subspaceColor}
        \hspace{-2px},
        \claimSpecColorNEW[Christopher Nolan's]{\claimVarStyle{$F_1$}}{\focusColor}
        films comprised
        \claimSpecColorNEW[34.8\%]{\claimVarStyle{V}}{\valueColor}
        of the total
        \claimSpecColorNEW[gross]{\claimVarStyle{M}}{\measureColor}
        for
        \claimSpecColorNEW[movies]{\claimVarStyle{ID\_key}}{\identifiedColor}
        with an
        \claimSpecColorNEW[IMDb score over 7]{\claimVarStyle{$S_2$}}{\subspaceColor}
        \hspace{-2px}.
        \claimExampleSpacing
        &
        \texttt{\{%
            "measure": \textcolor{\measureColor}{\claimVarStyleNEW{``gross''}},
            "value": \textcolor{\valueColor}{\claimVarStyleNEW{``34.8\%''}},
            "focus": [\textcolor{\focusColor}{\claimVarStyleNEW{\{"director" = "Christopher Nolan"\}}}],
            "subspace": [\textcolor{\subspaceColor}{\claimVarStyleNEW{\{"year" = 2013\}}}, \textcolor{\subspaceColor}{\claimVarStyleNEW{\{"IMDb\_score" > 7\}}}],
            "identifier\_key": \textcolor{\identifiedColor}{\claimVarStyleNEW{``movies''}}%
        \}}  
        \\ 
        \midrule         
       Trend 
        &
        \claimExampleSpacing
        From
        \claimSpecColorNEW[March 2020]{\claimVarStyle{$S_1$}}{\subspaceColor}
        to
        \claimSpecColorNEW[March 2021]{\claimVarStyle{$S_2$}}{\subspaceColor}
        \hspace{-2px},
        the
        \claimSpecColorNEW[number of COVID-19 cases]{\claimVarStyle{M}}{\measureColor}
        in the
        \claimSpecColorNEW[US]{\claimVarStyle{$S_3$}}{\subspaceColor}
        showed an
        \claimSpecColorNEW[increase]{\claimVarStyle{V}}{\valueColor}
        \hspace{-2px}.
        \claimExampleSpacing
        &
        \texttt{\{%
            "measure": \textcolor{\measureColor}{\claimVarStyleNEW{``case''}},
            "value": \textcolor{\valueColor}{\claimVarStyleNEW{``increase''}},
            "subspace": [%
                \textcolor{\subspaceColor}{\claimVarStyleNEW{\{"date" >= "March 2020"\}}},
                \textcolor{\subspaceColor}{\claimVarStyleNEW{\{"date" <= "March 2021"\}}},
                \textcolor{\subspaceColor}{\claimVarStyleNEW{\{"country" = "US"\}}}%
            ]%
        \}}
        \\    
        \midrule     
      Extreme
        &
        \claimExampleSpacing
        \claimSpecColorNEW[Glenlivet 18]{\claimVarStyle{F}}{\focusColor}
        has the
        \claimSpecColorNEW[highest]{\claimVarStyle{V}}{\valueColor}
        \claimSpecColorNEW[rating]{\claimVarStyle{M}}{\measureColor} 
        among
        \claimSpecColorNEW[whiskies]{\claimVarStyle{ID\_key}}{\identifiedColor}
        originating from
        \claimSpecColorNEW[Scotland]{\claimVarStyle{$S_1$}}{\subspaceColor}
        \hspace{-1px}.
        \claimExampleSpacing
        &
        \texttt{\{%
            "measure": \textcolor{\measureColor}{\claimVarStyleNEW{``rating''}}, 
            "value": \textcolor{\valueColor}{\claimVarStyleNEW{``max''}}, 
            "focus": [\textcolor{\focusColor}{\claimVarStyleNEW{\{"brand" = "Glenlivet 18"\}}}], 
            "subspace": [\textcolor{\subspaceColor}{\claimVarStyleNEW{\{"origin" = "Scotland"\}}}], 
            "identifier\_key": \textcolor{\identifiedColor}{\claimVarStyleNEW{``whiskies''}}%
        \}}
        \\
        \midrule 
      Rank 
        & 
        \claimExampleSpacing
        Among
        \claimSpecColorNEW[players]{\claimVarStyle{ID\_key}}{\identifiedColor}
        in
        \claimSpecColorNEW[point guards position]{\claimVarStyle{$S_1$}}{\subspaceColor}
        who played
        \claimSpecColorNEW[more than 60 games]{\claimVarStyle{$S_2$}}{\subspaceColor}
        in
        \claimSpecColorNEW[2023]{\claimVarStyle{$S_3$}}{\subspaceColor}
        \hspace{-3px},
        \claimSpecColorNEW[Trae Young]{\claimVarStyle{F}}{\focusColor} 
        is ranked
        \claimSpecColorNEW[4th]{\claimVarStyle{V}}{\valueColor}
        in
        \claimSpecColorNEW[three-point attempts]{\claimVarStyle{M}}{\measureColor}
        \hspace{-1px}.
        \claimExampleSpacing
        & 
        \texttt{\{%
            "measure": \textcolor{\measureColor}{\claimVarStyleNEW{``3PA''}}, 
            "value": \textcolor{\valueColor}{\claimVarStyleNEW{4}}, 
            "focus": [\textcolor{\focusColor}{\claimVarStyleNEW{\{"player"="Trae Young"\}}}], 
            "subspace": [%
                \textcolor{\subspaceColor}{\claimVarStyleNEW{\{"position" = "PG"\}}},
                \textcolor{\subspaceColor}{\claimVarStyleNEW{\{"games\_played" > 60\}}},
                \textcolor{\subspaceColor}{\claimVarStyleNEW{\{"year" = 2023\}}}%
            ], 
            "identifier\_key": \textcolor{\identifiedColor}{\claimVarStyleNEW{``players''}}%
        \}}
        \\     
        \midrule       
      Association 
        & 
        \claimExampleSpacing
        There's a
        \claimSpecColorNEW[positive correlation]{\claimVarStyle{V}}{\valueColor}
        between a
        \claimSpecColorNEW[movie's]{\claimVarStyle{ID\_key}}{\identifiedColor}
        \claimSpecColorNEW[budget]{\claimVarStyle{$M_x$}}{\measureColor}
        and its
        \claimSpecColorNEW[gross]{\claimVarStyle{$M_y$}}{\measureColor}
        earnings.
        \claimExampleSpacing
        & 
        \texttt{\{%
            "measure\_x": \textcolor{\measureColor}{\claimVarStyleNEW{``budget''}}, 
            "measure\_y": \textcolor{\measureColor}{\claimVarStyleNEW{``gross''}}, 
            "value": \textcolor{\valueColor}{\claimVarStyleNEW{``positive''}}, 
            "identifier\_key": \textcolor{\identifiedColor}{\claimVarStyleNEW{``movies''}}%
        \}}
        \\      
        \midrule
        Difference
        & 
        \claimExampleSpacing
       During the
        \claimSpecColorNEW[2019 NBA season]{\claimVarStyle{$S_1$}}{\subspaceColor}
        \hspace{-3px},
        \claimSpecColorNEW[James Harden]{\claimVarStyle{$F_x$}}{\focusColor}
        outscored
        \claimSpecColorNEW[Stephen Curry]{\claimVarStyle{$F_y$}}{\focusColor} 
        by 
        \claimSpecColorNEW[6.1]{\claimVarStyle{V}}{\valueColor} 
        \claimSpecColorNEW[points]{\claimVarStyle{M}}{\measureColor}
        \hspace{-1px}.
        \claimExampleSpacing
        & 
        \texttt{\{%
            "measure": \textcolor{\measureColor}{\claimVarStyleNEW{``points''}}, 
            "value": \textcolor{\valueColor}{\claimVarStyleNEW{6.1}}, 
            "focus\_x": \textcolor{\focusColor}{\claimVarStyleNEW{\{"player" = "James Harden"\}}}, 
            "focus\_y": \textcolor{\focusColor}{\claimVarStyleNEW{\{"player" = "Stephen Curry"\}}}, 
            "subspace": [\textcolor{\subspaceColor}{\claimVarStyleNEW{\{"season" = "2019"\}}}]%
        \}}
        \\
        \midrule 
        Categorization 
        & 
        \claimExampleSpacing
        There are
        \claimSpecColorNEW[seven]{\claimVarStyle{V}}{\valueColor}
        \claimSpecColorNEW[movies]{\claimVarStyle{ID\_key}}{\identifiedColor}
        that have an
        \claimSpecColorNEW[IMDb score over 9]{\claimVarStyle{$S_1$}}{\subspaceColor}
        and a
        \claimSpecColorNEW[gross of more than 300 million]{\claimVarStyle{$S_2$}}{\subspaceColor}
        \hspace{-1px}.
        \claimExampleSpacing
        & 
        \texttt{\{%
            "value": \textcolor{\valueColor}{\claimVarStyleNEW{7}}, 
            "subspace": [\textcolor{\subspaceColor}{\claimVarStyleNEW{\{"IMDb\_score" > 9\}}}, \textcolor{\subspaceColor}{\claimVarStyleNEW{\{"gross">"300,000,000"\}}}], 
            \texttt{"identifier\_key"}: \textcolor{\identifiedColor}{\claimVarStyleNEW{``movies''}}%
        \}}
        \\       
        \midrule       
       Distribution 
        & 
        \claimExampleSpacing
        The
        \claimSpecColorNEW[acceptance rates]{\claimVarStyle{M}}{\measureColor}
        of
        \claimSpecColorNEW[colleges]{\claimVarStyle{ID\_key}}{\identifiedColor}
        follow a
        \claimSpecColorNEW[right-skew distribution]{\claimVarStyle{V}}{\valueColor}
        \hspace{-1px}.
        \claimExampleSpacing
        & 
        \texttt{\{%
            "measure": \textcolor{\measureColor}{\claimVarStyleNEW{``acceptance rates''}}, 
            "value": \textcolor{\valueColor}{\claimVarStyleNEW{``right-skew distribution''}}, 
            "identifier\_key": \textcolor{\identifiedColor}{\claimVarStyleNEW{``colleges''}}%
        \}}
        \\   
        \midrule     
        Outlier 
        & 
        \claimExampleSpacing
        The
        \claimSpecColorNEW[movie]{\claimVarStyle{ID\_key}}{\identifiedColor}
        \claimSpecColorNEW[``Oppenheimer'']{\claimVarStyle{F}}{\focusColor}
        has a
        \claimSpecColorNEW[gross]{\claimVarStyle{M}}{\measureColor}
        that's quite the outlier among
        \claimSpecColorNEW[historical biopic]{\claimVarStyle{$S_1$}}{\subspaceColor}
        \hspace{-1px}.
        \claimExampleSpacing
        & 
        \texttt{\{%
            "measure": \textcolor{\measureColor}{\claimVarStyleNEW{``gross''}}, 
            "focus": \textcolor{\focusColor}{\claimVarStyleNEW{\{"movie = "Oppenheimer"\}}}, 
            "subspace": [\textcolor{\subspaceColor}{\claimVarStyleNEW{"genre" = "historical biopic"}}], 
            "identifier\_key": \textcolor{\identifiedColor}{\claimVarStyleNEW{``movies''}}%
        \}}
        \\   
        \bottomrule   
    \end{tabular}
\end{imageonly}
    \caption{Examples of the 10 data fact types along with the corresponding JSON output from our LLM pipeline.
    }
    \label{facttypes}
\end{table*}
    
\subsubsection*{\textbf{S1. Data Claim Detection.}} 
    This step corresponds to C1 in our framework. We use GPT-3.5 to classify tokenized sentences as either a `data claim' or not, guided by our task description and examples.

\subsubsection*{\textbf{S2 \& S3. Compound Claim Classification and Decomposition.}} \textcolor{black}{These two steps address the potential presence of compound claims~(F1). We first ask the model to distinguish if a data claim is a `single' or `compound' claim~(S2).} The compound claims are subsequently decomposed into distinct data facts while the single claims are kept intact, ultimately producing a list of decomposed data fact dictionaries~(S3). These dictionaries contain both the original sentence strings and the decomposed data fact strings.

\subsubsection*{\textbf{S4 \& S5. Coreference and Ellipsis Resolution.}}
\textcolor{black}{Addressing F3\&4,} for each data fact string, we instruct GPT to conduct \textcolor{black}{coreference resolution (\ie replacing pronouns with entities in the dataset) and ellipsis resolution (identifying and restoring omitted information, e.g., year=2023, based on the context of the input data document.}


%

\subsubsection*{\textbf{S6. Data Fact Classification.}} This step classifies each single claim as one of the ten data fact types. For the prompt, we provide a task description for classification along with our definitions of each data fact type, supplementing definitions with specific examples. 

\subsubsection*{\textbf{S7. Fact Specification Transformation.}} Next, we instruct GPT to convert the data fact strings into a type-specific JSON specification (\autoref{facttypes}). The JSON specifications are designed to capture the necessary key-value pairs for fact-checking claims that belong to the matching type. 
In this step, the attribute names of the reference dataset are provided in the prompt as contextual information to improve \sys{}'s ability to accurately parse the semantics.

\subsubsection{\rv{{\textbf{Veracity Determination}}}} 
Utilizing the JSON specifications derived from \textbf{S7}, \sys{} retrieves relevant data evidence from the provided reference dataset and \rv{computes precise values based on the specified data aggregation and operations. Specifically, for objective value-based fact types, including \textit{value, rank, proportion, extreme, difference}, and \textit{categorization}, \sys{} directly compares the \emph{claimed value} to the \emph{actual value}. For \textit{trend}, \sys{} is restricted to comparing only the two end values within a given timeframe. For other fact types (i.e., \textit{outlier, association, distribution}), \sys{} employs established mathematical calculations and rules to evaluate veracity. For \textit{outlier} detection, the interquartile range is applied to identify univariate outliers, while covariance matrix is used to detect bivariate outliers. The Pearson correlation coefficient is used to assess \textit{association}, and the skewness formula is applied to determine whether a \textit{distribution} is left- or right-skewed.}

\subsection{Designing \sys{}'s Interface}\label{visdesign}

Once the veracity is determined and the supporting data evidence is obtained, the results must be effectively communicated to the audience, substantiating the verdict with interpretable evidence and explanations, and empowering practitioners to act upon these insights. This section addresses this challenge by designing visual representations for demonstrating data evidence~(C5) and interactions to support actions from practitioners~(C6).



\begin{figure*}[t]
    \centering
    \includegraphics[width=1\textwidth]{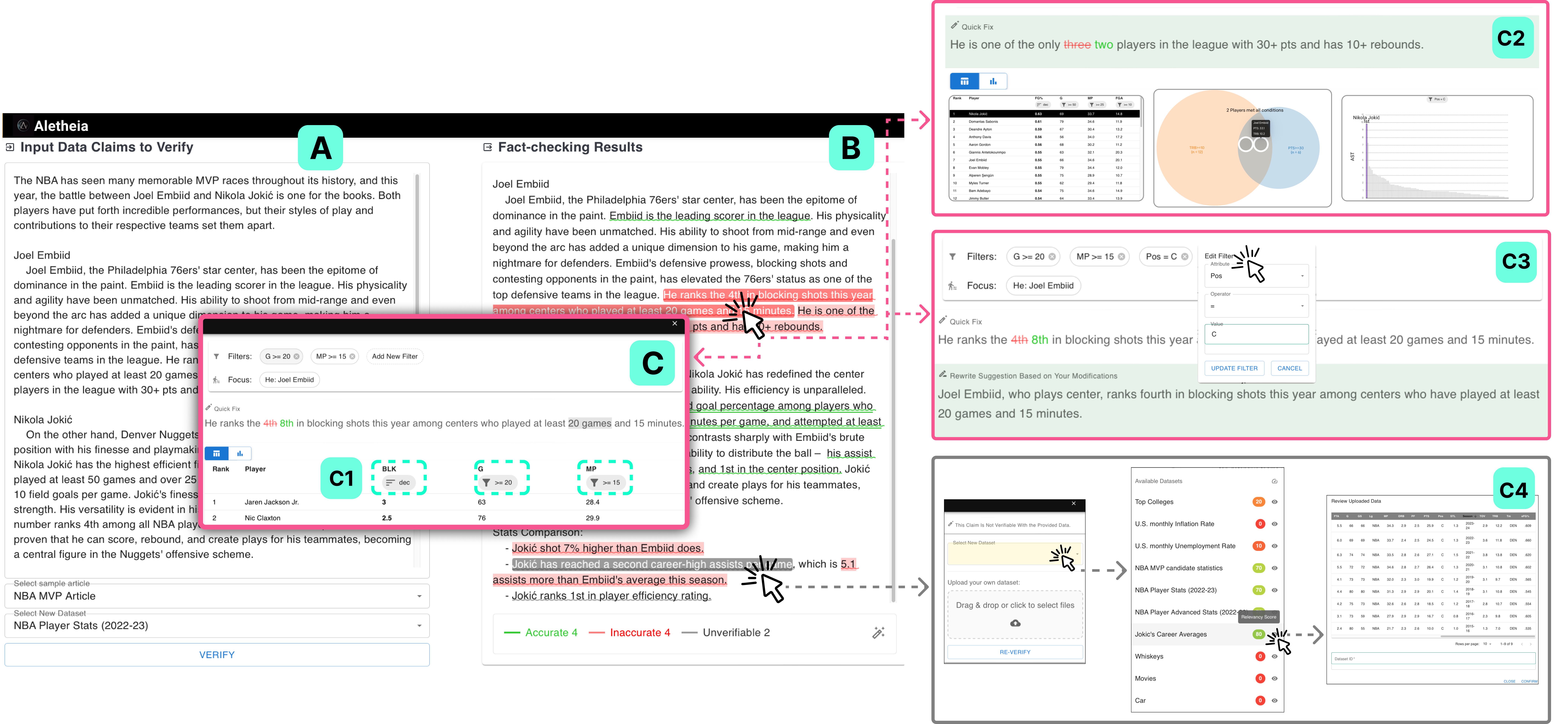}
    \caption{\sys's interface. Users enter textual content and select/upload a reference dataset in \textit{Input View (A)}. The backend then detects data claims, retrieves corresponding data evidence, and verifies them. The fact-checking results are presented in ~\textit{Result View (B)}, utilizing color codings to signify their verdicts: \textcolor{lightgreen}{\emph{accurate}}, \textcolor{lightred}{\emph{inaccurate}}, and \textcolor{gray}{\emph{unverifiable}}. Users click on the highlighted data claims to access the ~\textit{Evidence View (C)}. This view contains the designed data evidence presentation and interactions. 
    }
    \Description{}
    \label{fig:Aletheia}
\end{figure*}

\subsubsection{Designing Data Evidence Representations}

Existing research lacks comprehensive guidance on effectively presenting data evidence. Thus, our main objective is to explore the design space for representing data evidence. We specifically focus on two common approaches: data tables and visualization charts. Data tables, as a conventional and widely accessible form of data representation in the data fact-checking workflow, serve as a baseline. Visualization charts, known for their capacity to handle larger datasets, improve readability~\cite{joyce2019seeing}, \rv{and enhance human cognition through rapid perceptual inference and pattern recognition~\cite{Fekete2008}}, may offer a more efficient means of presenting data evidence. We propose designs for data table and visualization chart representations tailored to 13 subcategories derived from 10 data fact types: \textit{value (mean)}, \textit{value (median)}, \textit{value (sum)}, \textit{proportion}, \textit{trend}, \textit{extreme}, \textit{rank}, \textit{association}, \textit{difference}, \textit{categorization}, \textit{distribution}, \textit{1-D outlier}, and \textit{2-D outlier}.

To develop these designs for data fact-checking, we identified \rv{the following} three design goals. \rv{Our designs are also informed by literature on foundational visualization values~\cite{Fekete2008} and strategies counter cognitive biases (e.g. \cite{Holder_Xiong_2023_hiding_variablility}). We engaged in an iterative design process with active participation from two authors and feedback from a senior visualization researcher.} Details of our design choices and the corresponding illustrations for both the data tables and the visualizations are available in the \autoref{appendix:evidencepresentation}. Further design details and enlarged versions are available in the supplementary materials.


\subsubsection*{\textbf{DG1. Ensure alignment with the original data insights conveyed by the data claim.}} 
Each data fact type inherently provides a unique insight. For example, \emph{Rank} (\autoref{fig:visexample} V7) highlights an entity's position within a group. Effective verification of the accuracy necessitates that the evidence representation not only corresponds with the insight (e.g., the particular rank, \textit{`8th'}) but also encapsulates the scope of the relevant data (e.g., a sorted list that includes the rank). Similarly, a \emph{Categorization} fact (\autoref{fig:visexample} V10) indicates an entity's affiliation with a subset of entities. Thus, the design should display the entity, the intersection, and the inclusion criteria.    
The variety of insight types emphasizes the need for customized representations tailored to specific data fact types or subtypes. 

\subsubsection*{\textbf{DG2. Streamline viewers' analysis of data evidence, facilitating quicker judgments.}}
\textcolor{black}{Efficiency is key in designing data evidence presentations for fact-checking, as it could enhance professionals' productivity and increase general readers' adoption.} 
To achieve this goal, we embraced three key design approaches. First, we display only the relevant data segments, eliminating the need to sift through the full dataset. Second, we highlight and annotate salient elements, especially data points mentioned in the claim. Third, for claims involving derived statistics, we automate data operations and computation to directly show the summary statistics.

\subsubsection*{\textbf{DG3. Bolster viewers' confidence in their assessments.}} Another critical element we emphasize is viewers' confidence in their judgment. Given that both the verdict and the representation provided are rooted in the same data evidence, we posit that a representation that boosts viewers' confidence can also enhance their trust in the system's ability to retrieve accurate evidence and process it appropriately. We first ensure the transparency of data operations either by incorporating \emph{operation widgets} (\autoref{fig:Aletheia} C1) or directly through visual encodings. Next, we emphasize the presentation of individual data points, aiming to provide an overview of the data distribution that allows for a `sanity check.'


\subsubsection{Interface Design} 
In addition to designing evidence representations, we explore the feasibility of integrating them into an interactive application along with our proposed fact-checking framework. We envision a scenario where authors must verify and rectify inaccurate data claims, and thus propose three key design requirements:

\subsubsection*{\textbf{DR1. Facilitate rapid reviewing and correction of erroneous claims.}}
Upon receiving fact-checking results, authors naturally want to review the verdicts and supporting data evidence. This review allows them to decide whether to trust the results and take action, e.g., to correct inaccuracies in the data claim. Thus, an interactive system should streamline the review and revision process.


\subsubsection*{\textbf{DR2. Enable user intervention for AI mistakes.}} AI mistakes may occur when GPT incorrectly infers the data subspace or coreferences (S4\&5), especially when keywords are missing or ambiguous in the data claim. Also, GPT may incorrectly parse a focused attribute (S7). Under these circumstances, it is critical to incorporate human supervision and intervention for enhanced reliability.



   
\subsubsection*{\textbf{DR3. Support integration of additional reference data.}} Real-world articles can simultaneously rely on multiple datasets with varying contexts, thereby posing a challenge when attempting to fully automate the fact-checking process. 
Enabling users to add additional reference datasets to evaluate claims that are initially unverifiable could enhance \sys's utility. Furthermore, considering users' potential unfamiliarity with the `suitability '~\cite {vlachos-riedel-2014-fact} of reference data, the system should provide support for users to assess whether a dataset is appropriate for verifying certain data claims.



\subsection{\sys{} Workflow and Usage Scenario}
We integrated our LLM-based pipeline with the interactive user interface, resulting in a functional prototype, \sys. 
The backend of \sys{} is built on Python Flask, leveraging OpenAI (GPT-4) for its NLP tasks. The frontend is developed using React, with the visualizations implemented in D3.js.  
As shown in \autoref{fig:Aletheia},~\sys's interface has three main views: an input view~(A), a result view~(B), and an evidence view~(C). 
To illustrate the utility of \sys{}, consider the scenario of a sports editor, Jordan, tasked with reviewing the accuracy of a written article containing various data claims. 

Jordan first loads the draft article of NBA MVPs into \sys{}'s \textit{Input View}, and selects a reference dataset of players' average statistics from \textit{Basketball-Reference}, an authoritative sports data platform. Upon requesting \sys{} to verify the claims, the \textit{Result View} generates a fact-checking report with the data claims color-coded. Jordan navigates through these data claims and toggles between \emph {table} and \emph{visualization} forms to examine the data evidence. This interaction helps him further assess the verification results and determine if he should apply a quick correction suggested by \sys. For instance, as shown in \autoref{fig:Aletheia} C2, \sys{} recommends correcting a \emph{ranking value} error from \textit{`4th'} to \textit{`8th'} based on the computed results. \sys{} also supports a `quick rectify' action, which can be useful when many instances of text-data misalignment occur (e.g., during data updates). These system functions support~\textbf{DR1}.

To mitigate the risks associated with inferential mistakes~(\textbf{DR2}), \sys{} provides an interactive widget (\autoref{fig:Aletheia}~C3). This widget displays the key AI inferences, with filters for tuning the subspace, coreferences, and focused attributes. Jordan hovers through these `chips' to examine associated text segments, and identifies that the AI overlooked a filter (Position=Center), which should have been applied to the phrase `among all centers' in the claim. \sys{} enables him to directly edit these `chips' to override the AI's inferred elements, \ie to add missing filters. This action prompts \sys{} to reassess the associated claim based on Jordan's modifications and simultaneously propose a text revision reflecting these adjustments. 

After addressing the discrepancies flagged in red, Jordan encounters two claims marked as `unverifiable' (\autoref{fig:Aletheia} C4). To resolve these, he imports new datasets for targeted evaluation. These extra datasets are bound to individual claims and do not impact other verified claims. Employing \sys's \textit{data relevance evaluation} function, which leverages the GPT model to assign a `suitability score' based on the attribute names and the claim, Jordan can quickly compare and identify the most pertinent datasets~(\textbf{DR3}). This prompts \sys{} to reassess the accuracy of the targeted claim. 

\section{Prompting Framework Evaluation}
\label{section5}
%

In this section, we evaluate the feasibility of \sys's LLM-based pipeline, with a particular focus on two core steps: 
\textbf{data fact classification} (\emph{S6}) and \textbf{data facts specification transformation} (\emph{S7}) in \autoref{fig:prompt_pipeline}.
\textcolor{black}{We focus on these core steps for three reasons. First, conducting a comprehensive evaluation requires viable testing datasets of data documents with corresponding reference datasets, which are currently unavailable and expensive to curate. 
Second, our initial experiments with real-world data articles (\eg \cite{levy_2023, msn_2023}) indicate that GPT-3.5 is competent in identifying data claims from passages/articles; our unoptimized prompt achieved an accuracy of 87.2\% and 93.1\%, respectively (see the supplementary materials).
Third, although we utilize GPT to perform the pre-processing steps, the downstream tasks (e.g., coreference, ellipsis resolution) have been extensively explored in NLP research, with pre-trained statistical models showing increasing capabilities~\cite{aralikatte-etal-2021-ellipsis, joshi-etal-2020-spanbert}.} 

\subsection{Data Curation}
Existing benchmarking datasets for automated fact-checking (e.g., FEVER~\cite{thorne-etal-2018-fever}, LIAR~\cite{wang_2017_liar}, MultiFC~\cite{Augenstein_2019_multiFC}, ClaimBuster~\cite{fatma_arslan_2020_3836810}, etc.) primarily focus on text-sourced (i.e., knowledge-based) claims. For example, LIAR~\cite{wang_2017_liar}, based on human-labeled shorts claims from PolitiFact~\cite{politifact}, includes statements like \say{\emph{Newly elected Republican senators sign pledge to eliminate food stamp program in 2015.}} These text-sourced datasets do not align with the focus of this work, and there is a notable absence of open-sourced benchmarking datasets tailored for data claims. In the limited body of work specifically addressing data/statistical claims (e.g., \cite{Karagiannis_2020_Scrutinizer, rezgui2021automatic}), training/testing datasets are often synthesized with templates due to the scarcity of benchmarking datasets. In this work, we employ a similar template-based data curation approach, focusing more on diverse types of data insights~\cite{Srinivasan_2019_Voder, Wang_2020DataShot, Shi_2021_Colliope} to cover a range of insight categories. 

We programmatically curated ground truth claims for 10 aggregated data fact types, generating 40 template-based claims per type along with their corresponding data fact JSON specifications, following \autoref{facttypes}.
To better represent the language variation in real-world claims, we employed GPT to produce a paraphrased version of each claim. We compiled a dataset of 400 test claims, each featuring data fact specifications, a claim generated from type-specific templates, and a paraphrased version. We manually reviewed the paraphrased claims to ensure they preserved the original data facts.


\subsection{Evaluation Results}

\subsubsection*{Data Fact Classification}\label{sec:data-fact-classification}
\rv{We tested our data fact type classifier on a balanced dataset of 400 paraphrased examples using GPT-4. GPT-4 achieved perfect classification against natural language variation.} The results indicate that \textbf{\textit{an LLM can robustly classify data facts into types following the data fact taxonomy},} simplifying previous approaches that required multiple algorithms~\cite{walenz2014finding}.

\subsubsection*{Fact Specification Transformation}\label{eval:transformation} 
\begin{figure}[t]
    \centering  
    \includegraphics[width=1\columnwidth]{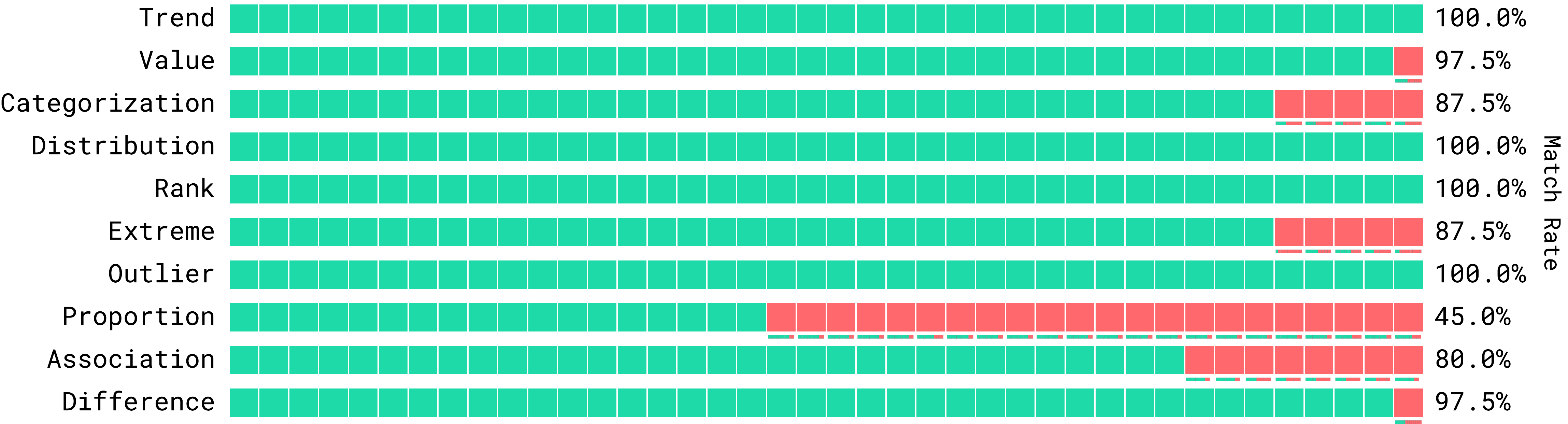} 
    \caption{Success rate of data fact specification transformation. Each row corresponds to a distinct data fact type. Boxes within the rows represent individual examples. Green boxes indicate successful transformations, where all transformed attributes and values match the ground truth. Red boxes represent examples with incomplete or incorrect conversions. The small rectangles below the red boxes represent partial match performance with the same color code. The average rate of complete matches is \textcolor{black}{89.5\%}. 
    }
    \Description{Data Fact Specification Transformation Success Rate}
    \label{fig:spec_transformaion} 
\end{figure}

Next, we test our fact specification transformation step, taking the preprocessed, GPT-paraphrased claims as input and outputting a JSON specification for each claim. We present matching accuracy for each data fact type in \autoref{fig:spec_transformaion}, with the green area representing complete matches (i.e., the generated JSON \textit{exactly} matches the ground truth) and the red area the partial matches (i.e., \textit{parts} of the generated JSON match those in the ground truth). There are no cases in which the ground truth is entirely missed, and the ratio of partial matches is also represented below each red box.

Across all data fact types, the average rate of complete matches is 89.5\% ($\sigma$ = 16.27), which means that nearly 90\% of the generated JSONs are fully usable for data evidence retrieval. For comparison, state-of-the-art LLMs for code generation, when given a single trial, produce executable code around 20-40\% of the time (see ``Pass Rate'' or ``Pass@1'' in  \cite{li2023starcoder, chen2021evaluating}), showing how much more challenging the generation of executable representations is when compared to classification tasks such as in \autoref{sec:data-fact-classification}. We consider the current average rate of complete matches (89.5\%) positive evidence that \textbf{\textit{an LLM can be used to generate well-formed data fact specifications}}, allowing for automated systems such as \sys. 

\subsection{\textcolor{black}{Failure Case Analysis}}
Nevertheless, there is more room for improvement in fact specification transformation: the fact type ``Proportion'' has the lowest rates (45\%) of complete matches. The primary cause of failure can be traced back to the parsing of data filters --- achieving a complete match in the subspace requires identifying all applied filters. \rv{Consider a \emph{Proportion} data claim: ``\textit{In 2013, 75.59\% of the budget for Italian movies was spent on movies with an IMDB score higher than 6.}'' A full match in parsing requires the detection of a focus space filter \textit{(IMDB\_score $\geq$ 6)} and two subspace filters \textit{(released\_year = 2013 and country = Italy)}.
This complexity explains why \emph{Proportion} type has the lowest \textit{complete matching} score --- it requires not only the identification of all filters but also a clear distinction between \emph{focus\_space} and \emph{subspace}.} This bottleneck could justify adding new steps to our pipeline to better perform this task for this specific set of fact types. \rv{It is important to note that data claims can become more complex regarding fact type, semantics, and compoundness, posing greater challenges. We further discuss \sys's limitations and future opportunities in \autoref{sec:futureresearch}.}

\section{User Evaluation of Data Evidence Representations (Charts \& Tables)}\label{userstudy}
We conducted a mixed-method user study with 20 participants, using our 26 data evidence representations (\autoref{visdesign}) as probes to gather quantitative behavioral insights and qualitative feedback. 


\subsubsection*{Participants Demographics}
We recruited a total of 20 participants from our institution for the study, comprising 12 males~(60\%) and 8 females~(40\%). The majority, 18 participants, are aged between 25 and 34, with 2 aged 18 to 24. All are either graduate degree holders or candidates, with diverse backgrounds in statistics, data analysis, data visualization, and varying experiences with data articles.

\subsubsection*{Data Curation}
To maintain a consistent and reliable study environment, we chose a manually curated, predetermined test dataset over real-time API calls to GPT, due to the potential for the model's variable runtime to disrupt controlled conditions. We curated four data claims --- two accurate and two inaccurate ones --- for each of the 13 data fact types. To reduce participant workload, we split the 52 test data claims into two datasets. Each dataset contains an accurate and an inaccurate claim for every fact type, resulting in 26 tasks per participant per study phase (2*13).


\subsubsection*{User Study Interface} 

Given the interactive features of our data evidence representations and the need to monitor assessment duration, we devised a study-tailored interface (\autoref{fig:userinterface}) based on \sys
. This study interface consists of five components: a tutorial page (C), two study session pages (A \& B), a page for visualizing participants' results (D), and an ``Exploration Page (E)''  that helps participants review the representations and answer interview questions.

\subsection{User Study Procedure}
Our study consists of three phases. \emph{Phase I} and \emph{Phase II} focus on collecting quantitative data on \emph{assessment time} (tracking), \emph{confidence shift} (self-reporting), and \emph{preference} (self-reporting). \emph{Phase III} is a post-study interview. A detailed interview protocol is available in the supplemental material. The entire study process was video-recorded, and the audio was transcribed for qualitative analysis. 

\subsubsection*{\textbf{Phase I}} Participants were randomly given one of the two datasets in a counter-balanced manner and assigned to either the \textit{Table Group} (Group A) or the \textit{Visualization Group} (Group B). Each group used the respective data evidence representations to review the data claims. Prior to assessing the claims, participants underwent a five-minute tutorial session to familiarize themselves with the data encodings for various data fact types. We also showcased the interface for Phase I using a demo dataset. 

During the main part of \emph{Phase I} (shown in \autoref{fig:userinterface} (A)), participants were sequentially presented with individual claims. They were instructed to read the claim carefully and then click the ``Verify'' button to view the model's verdict. The system then fetched the corresponding data evidence and displayed a verdict stating ``\textit{The AI determines this claim to be accurate/inaccurate.}'' Participants were informed beforehand that the \textit {``AI prediction''} might be inaccurate and should only be considered as a reference. After viewing the verdict, participants were instructed to assess the veracity of the claim based on the data evidence revealed after clicking the ``Show Data Evidence'' button. A timer started upon the appearance of the visual representations. After reviewing the data evidence, participants indicated their decision by selecting either ``Accurate'' or ``Inaccurate'', which simultaneously stopped the timer. Participants also self-reported their confidence level on a scale from 1 to 5.

\subsubsection*{\textbf{Phase II}} At the beginning of \emph{Phase II}, participants spent five minutes familiarizing themselves with the counterpart representations (i.e., participants in the Table Group were presented with visualizations in this phase and vice versa). We again showcased the interface for Phase II using a demo dataset. Participants performed the same fact-checking task as in Phase I, but with counterpart representations. After making a judgment, participants' confidence level from Phase I was disclosed, along with the corresponding Phase I representations (as shown in \autoref{fig:userinterface} (B)). Participants were instructed to compare their confidence with their confidence Phase I ratings. Additionally, we asked participants to self-report their preference between the table and visualization representation for fact-checking the current claim using a five-point scale.

\subsubsection*{\textbf{Phase III}} Concluding the study, we presented a summarized visualization illustrating the participant's confidence shift and preferences between the two study parts~(\autoref{fig:userinterface} (D)), followed by a semi-structured interview to gather insights about their thinking process, perceived advantages/disadvantages, and feedback on both representation methods. During the interview, participants could use the ``Exploration Page''~(\autoref{fig:userinterface} (E)) to navigate the reviewed claims and the two corresponding visual representations while answering questions.

    
    
    
    

\subsection{Quantitative results} 
We \rv{report four measurements}: \textbf{(A)~\emph{time-spent}} assessing the data evidence in \emph{Phase I}, \textbf{(B)~\emph{confidence shift}} between the data evidence representation forms in \emph{Phase I} and \emph{Phase II}, \textbf{(C)~\emph{preference}} and \textbf{\rv{(D)~\emph{accuracy}}} between the two data evidence representations. The results are shown in \autoref{fig:timespent}. 

\subsubsection*{\textbf{Time-spent}}
\begin{figure*}[t]
    \centering
    \includegraphics[width=1\textwidth]{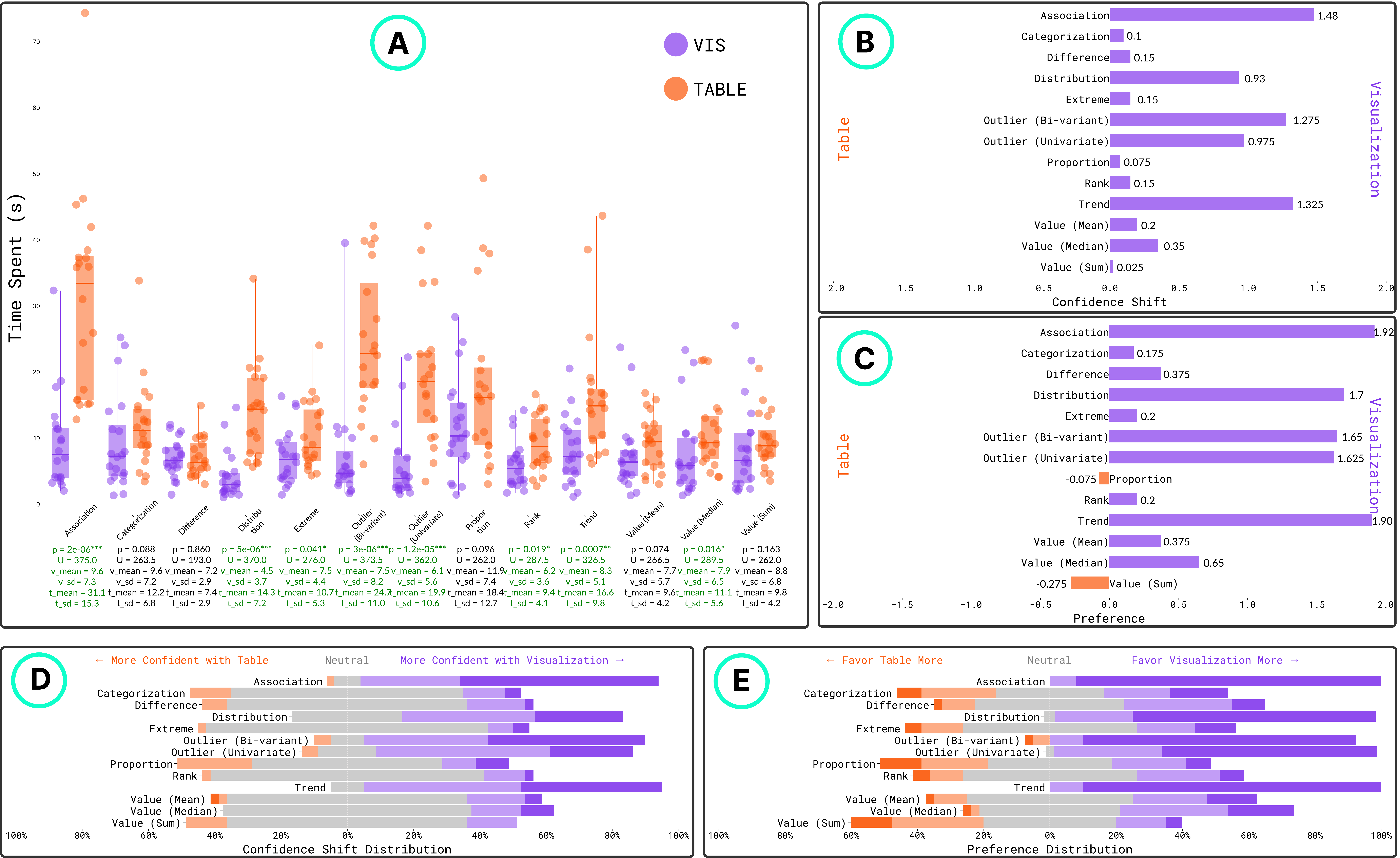}
    \caption{Quantitative results from our user study with 20 participants comparing visualizations and tables as different data evidence presentation forms. (A)~The distribution (box) and individual (point) time taken to assess the accuracy of thirteen distinct data facts. The x-axis represents the data fact types, while the y-axis indicates the duration in seconds. The two diverging bar charts show the average shift in (B)~the viewers' confidence and (C)~their preferences across the thirteen data fact types. \textcolor{mypurple}{Right-pointing bars} signify that participants have greater confidence in their assessment when using the \textcolor{mypurple}{visualization}, or they prefer to use visualizations for fact-checking the respective data facts. Conversely, \textcolor{myred}{left-pointing bars} indicate greater confidence or preference for \textcolor{myred}{tables}. \textcolor{black}{Figures (D) and (E) display the percentage distribution for each response option regarding confidence shift (D) and preference (E). The length of the bars represents the percentage of each selection on the five-point scale. \textcolor{gray}{Gray bars} represent \emph{Neutral}. \textcolor{myred}{Orange bars} represent \textcolor{myred}{\emph{Table}} while \textcolor{mypurple}{purple bars} represent \textcolor{mypurple}{\emph{Visualization}}. Darker \textcolor{mydarkred}{\textbf{red}} and \textcolor{mydarkpurple}{\textbf{purple}} signify greater intensity (i.e., much more confident/strongly favor). }
    }  
    \Description{}
    \label{fig:timespent}
\end{figure*}

We applied the Mann-Whitney U test at a significance level of 0.05 to determine whether there existed any statistically significant differences between the \emph{table} and \emph{vis} representation types across various data fact types. In general, participants who used visualization charts to assess the claim spent less time (\(M = 7.9, SD = 6.1\)) than those who used data tables (\(M = 15.0, SD = 10.7\)) in \emph{Phase I}. This time efficiency was consistent across all fact types, with statistical significance identified in the majority of fact types (8 out of 13), including \emph{Distribution} (\(U = 370.0, p < 0.001\)), \emph{Trend} (\(U = 326.5, p < 0.001\)), \emph{Rank} (\(U = 287.5, p < 0.03\)), \emph{Association} (\(U = 375.0, p < 0.001\)), \emph{Outlier (Univariate)} (\(U = 362.0, p < 0.001\)), \emph{Outlier (Bi-variant)} (\(U = 373.5, p < 0.001\)), \emph{Value (Median)} (\(U = 289.5, p < 0.03\)), and \emph{Extreme} (\(U = 276.0, p < 0.05\)). No significant differences were observed in the rest of the data fact types. More detailed mean, U-value, and p-values for each data fact type under two representations are reported in \autoref{fig:timespent}, with the statistically significant conditions marked in green.

\subsubsection*{\textbf{Average confidence shift}} Confidence shift is quantified using a five-point scale: -2 (much more confident with table), -1 (more confident with table), 0 (about the same), 1 (more confident with visualization), and 2 (much more confident with visualization). Positive values indicate higher confidence on average with visualizations for fact-checking, whereas negative values suggest the opposite. In general, \emph{vis} has slight advantage over \emph{table} in enhancing user confidence (\(M = 0.56, SD = 0.87)\). While this advantage is consistently observed across all fact types, it is most pronounced (\(M > 1\)) for \emph{Association} (\(M = 1.48, SD = 0.74\)), \emph{Trend} (\(M = 1.33, SD = 0.66\)), and \emph{Outlier (Bi-variant)} (\(M = 1.28, SD = 0.85\)). On the contrary, the least advantage was observed in types including \emph{Value (Sum)}, \emph{Proportion}, \emph{Categorization}, \emph{Difference}, \emph{Rank}, \emph{Extreme}, and \emph{Value (Mean)} with average confidence shifts less than 0.2.

\subsubsection*{\textbf{Preference}} Preference is measured using the same five-point scale: [-2 (strongly favor table), -1 (favor table), 0 (neutral), 1 (favor visualization), 2 (strongly favor visualization)]. Positive values indicate that participants prefer visualizations over tables for specific fact types, whereas negative values suggest a preference for tables. Overall, participants showed preference on \emph{vis} over \emph{table} (\(M = 0.81, SD = 1.17\)). This observation is consistent for the majority of the data fact types (11 out of 13), except \emph{Proportion} (\(M = -0.08, SD = 1.12\)) and \emph{Value (Sum)} (\(M = -0.28, SD = 1.04\)). Notably, participants exhibited the most pronounced preference on \emph{vis} for fact types including \emph{Association}, \emph{Trend}, \emph{Outlier (Univariate)}, \emph{Outlier (Univariate)}, \emph{Outlier (Bi-variant)} and \emph{Distribution} with an average preference score over 1.7. 

\subsubsection*{\rv{\textbf{Accuracy}}}
\rv{Both groups exhibit high accuracy when determining the veracity of the data statements. The overall accuracy (under time pressure) is 90.19\% ($\sigma = 0.3$). The table group achieved 89.62\% accuracy, while the visualization group achieved 90.77\%.   
The accuracy was higher for objective value-based claims, including \textit{Rank (100\%), Proportion (100\%), Difference (97.5\%), Categorization (95\%), Value~(Mean) (97.5\%), Value~(Median) (95\%), Value~(Sum) (100\%), Trend (87.5\%), Extreme (100\%)}, and lower for subjective ones, including \textit{Outlier (Univariate) (67.5\%), Outlier (Bivariate) (65\%), Association (78\%), Distribution (93.3\%)}. While the overall difference in detection accuracy is not significant, the visualization group achieved notably better accuracy in \textit{Association} (88\% vs. 68\%) but lower accuracy in \textit{Outlier} facts (60\% vs. 72.5\%). The lower detection accuracy in \textit{Outlier} facts can be attributed to the subjective nature of visual interpretation and varying algorithms and thresholds used to determine outliers. Nevertheless, the overall high detection accuracy indicates that our designed representations can effectively assist users in identifying inaccurate claims under time pressure.}

\subsection{Findings and Takeaways}

\subsubsection*{\textbf{T1. Visualizations inherently offer advantages when fact-checking data claims related to patterns or distributions across numerous data points.}}
Visualizations exhibit pronounced advantages over tables when verifying \emph{association}, \emph{distribution}, \emph{outlier}, and \emph{trend} data fact types, which is evident across all measurements: time-spent (\autoref{fig:timespent}A), participants' confidence~(\autoref{fig:timespent}B) and preference~(\autoref{fig:timespent}C).
Participants commonly expressed that visualizations are significantly more helpful for determining the veracity of data claims that necessitate an `overview of the data.' Particularly, P12 stated that \say{\textit{Visualizations is just a lot clearer than looking at a data table, especially if the table has a lot of rows you have to scroll through and process all of those information}}.    
This advantage can become more prominent for bigger datasets. Interestingly, \emph{extreme}, \emph{rank}, and \emph{value (median)} also exhibited statistically significant time improvements in visualizations over tables.
Given that statistics are readily available in the data table, verifying these claims only requires participants to process a single number. However, statistical significance was not found in \textit{categorization}, \textit{difference}, \textit{proportion}, and \textit{value (mean\&sum)}, five data types with similar settings. 
We anticipated that the visualization representation of the ranked bar chart for these three data types, offering visual confirmation of accurate sorting, could potentially reduce the time needed for viewers to be persuaded.

\subsubsection*{\textbf{T2. Displaying data operation widgets accelerates assessment and boosts confidence}}
We included operation widgets (\autoref{fig:Aletheia} C1) to indicate data operations in both \emph{table} and \emph{visualization} representations. 
19 participants agreed that these widgets bolstered their confidence in the system retrieving the correct data evidence. 
Participants noted that their initial mental task upon seeing the data evidence was to align the keywords with the filtering widgets. P6 stated that \say{\textit{it [showing widgets] is big, because I need to know, especially when a data claim is being made over a subset}}. P4 concurred that \say{\textit{these [filter widgets] are the ones that really affect my confidence\ldots it gives me an understanding of what subset of data the person was trying to analyze.}} Seeing these widgets also expedited their assessment, as it obviated the need to scrutinize individual values to confirm their presence in the subspace. In particular, P20 emphasized that \say{\textit{[not showing filters] would impact visualization a lot more}} because \say{\textit{[in table], I can quickly see it [the relevant information]}}. Though only occasionally checked the widgets, P10 recognized the importance of them:\say{\textit{[knowing] what they present matters a lot, and I can take a look any time I want.}} 
However, there was one outlier (P8), who assumed the data evidence was correctly retrieved, therefore, examining the widgets \say{\textit{made me spend a little bit more time and had no effect on my confidence.}} P2 also indicated that once they established trust in \sys's ability to apply the correct filters, they paid less attention to the widgets.

\subsubsection*{\textbf{T3. Unit representations enhance confidence, even when not scrutinized}}
During the interview phase, participants reflected on their attention to unit representations and how their absence might affect their confidence regarding fact types linked to specific summary statistics. While all participants agreed that they did not scrutinize the visual elements representing individual values, 15 out of 20 participants agreed that displaying summary statistics alone would reduce their confidence compared to pairing them with unit representations because the unit representations allow them to verify that the data distributions align with the aggregated statistics. For example, P2 stated that \say{\textit{I want to see raw data to make sure that the thing that I'm consuming is accurate.}} P7 elaborated on verifying computed values based on the underlying raw data: \say{\textit{This is one way of confirming that the average is correct\ldots~if the [average] bar is somewhere in the middle.}} P17 expressed that \say{\textit{without any individuals, you don't have a global understanding about the data points distribution.}} Three participants (P1, P5, P18) emphasized the need for visualizations where the distribution of units offers greater value for a "sanity-check'' of the provided statistics than mere ``numbers'' in the table.

\subsubsection*{\textbf{T4. Contextual information can be both reassuring and distracting in customized visualization}}
For the data fact types requiring only single summary statistics to verify the claim, we provide contextual information using different visual techniques. Three particular visuals --- \emph{value (sum)}, \emph{proportion}, and \emph{categorization} --- are particularly customized, but received low preference on average. A common reason cited by our participants during the interview phase was that these fact types solely require aggregated statistics to assess their veracity after confirming the subspace and focus, and it can take additional effort to process the additional contextual information. Participants also mentioned their struggle with unfamiliar chart types, leading to slower comprehension. For example, P10 explained that \say{\textit{\ldots~it took me more time to understand the mapping\ldots~so I feel a little bit distracted when trying to extract useful, relevant information to do the fact-check.}} Participants also appreciated the assurance of contextual information provided. P10 particularly liked the proportional Venn diagram, \say{\textit{it's just nicer\ldots~you get more information\ldots~for fact-checking, the context helps because it gives you an assurance that the data is valid and there's no arbitrary thing.}} P15 pointed out that, for \emph{sum values}, \say{\textit{using visualization, there is no concern about the total [sum operation] because the height should be the [total] height of each one.}}

\subsubsection*{\textbf{T5. Highlighting salient information streamlines the fact-checking process}} 
All participants concurred that a crucial mental step in fact-checking data claims involves extracting salient information from both the text and data evidence and then verifying their alignment. Participants also acknowledged that our design decision to \textit{highlight and annotate salient visual elements} aids in accelerating this process. Participants who preferred tables over visualizations for ``one-number'' fact-checking noted that tables provided them a clear location to focus on, typically the last sticky row we highlighted in the table. P5 expressed that\say{\textit{I know where to expect to see it,}}. However, it is not as consistent with visualizations, even though the visualizations included annotated labels with the same information. Participants in favor of the visualizations emphasized the intrinsic value of visualization as a form of abstract information. For example, P7 stated that the \say{\textit{it [visualization]
just abstracts away all the information that I don't need to know.}} P17 added that \say{\textit{abstracted information that corresponds to the statement is way more efficient}} when it comes to fact-checking.

\subsubsection*{\textbf{Design Recommendations for Presenting Data Evidence}} 

Drawing on our quantitative results, qualitative findings, and reflections on design, we derive four general design recommendations:

\begin{enumerate}[leftmargin=*]
    \item \textbf{Display data operations, especially the filters}: Regardless of the presentation format, our findings suggest the importance of consistently displaying the data operations executed to retrieve relevant data evidence and calculate aggregated statistics, particularly the filters used to determine the subspace (\textbf{T2}). Such transparency enhances viewers' trust in data validity.
    \item \textbf{Make salient information visually predominant}: After viewers establish trust in data validity, they search for the \emph{salient} information crucial to determining veracity (observed in \textbf{T5}). This salient information typically includes elements like \emph{value}, \emph{measure}, and \emph{focus} depending on the fact type. Making these elements visually predominant in the evidence presentation will streamline the verification process.
    \item \textbf{Key information first, contextual information on-demand}: Contextual information can enhance understanding and bolster trust, but an excess may overwhelm viewers since context is not essential for assessing fact-level veracity (as suggested in \textbf{T4}). Therefore, we recommend abstracting such information and enabling viewers to access it on demand, \ie via interactions.  
    \item \textbf{Enhance readability with visual aids}: While \sys{} offers computed statistics, comparing raw numbers is challenging when the values are large (e.g., movie grosses). As favored by participants (\textbf{T5}), we suggest including visual aids in visualization, like an additional line indicating the ``claimed value'', to ease comparison. \textcolor{black}{Visual aids (e.g., colored underlines) can also be used to highlight the mapping of entities, attributes, and filters between textual references in claims and visual counterparts, facilitating more efficient ``sanity check''.}   
\end{enumerate}

\section{Application Scenarios for \sys}\label{sec:application}
We design \sys{} specifically to assist authors and editors in the task of ensuring accuracy within data-rich articles. Given that many news organizations have already established their data infrastructure, inserting a system like \sys{} in their editorial workflow could markedly enhance their efficiency. Additionally, the methodologies underpinning \sys{} can be seamlessly integrated with data article authoring tools, such as CrossData~\cite{Chen_2022_CrossData}, DataTales~\cite{Sultanum_2023_DataTales}, reinforcing the accuracy of content produced. We envision \sys{}'s adaptability across diverse user groups, with potential applications ranging from browser extensions with reference data plugins for general news readers to fact-checking social media posts, financial reports and data-driven claims in other domains. Given these diverse applications, we propose the following two recommendations for tailoring \sys{} to specific domains or contexts:

\subsubsection*{\textbf{Leverage in-context examples in LLM prompts to enhance NLP task performance and accommodate domain-specific requirements}} We provide in-context examples in prompts for multiple NLP tasks in \sys{}. The LLM's flexibility enables easy customization of the pipeline to meet domain-specific requirements through mere modifications of the in-context examples and their associated definition and reasoning. Consider the case of a data fact not present in the data fact taxonomy, such as the \emph{``prominent streak''}~\cite{Jiang_2011_prominentstreak}. If this type of data claim is prevalent in a domain (e.g., sports), one can integrate its definition, examples, and desired output specifications in the chained prompts, thereby enabling LLM to retrieve the corresponding data evidence.

\subsubsection*{\textbf{Tailor data evidence representations for specific user groups, associated tasks, and domain conventions}} 
\begin{figure}[t]
    \centering
    \includegraphics[width=1\columnwidth]{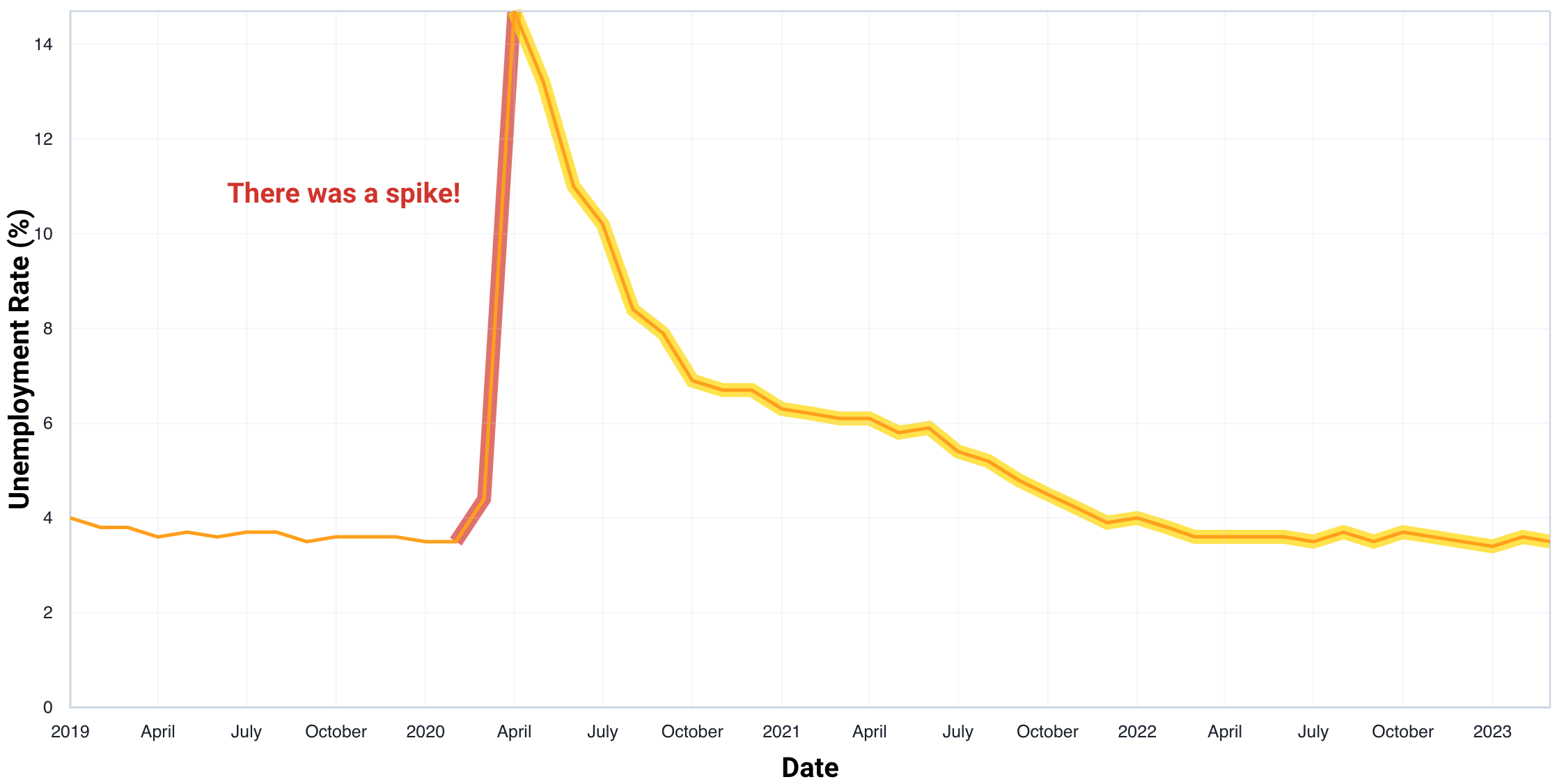}
    \caption{An example of showing contextual information. The data claim is ``\textit{The unemployment rate experienced a decrease between April 2020 and March 2023.}'' While the claim is technically accurate, it might be suspected of ``cherry-picking the timeframe''. Displaying the context can help inform audiences about the potential `pitfall.'}
    \Description{ }
    \label{fig:contextvis}
\end{figure}

While our 13 pairs of type-specific data evidence representations cover common data fact types, they are not exhaustive. Additional data fact types will require customized evidence presentations. For instance, a sequential bar chart that highlights instances meeting (or not meeting) specific conditions would be appropriate for representing a \emph{prominent streak} data fact.  
Furthermore, we recommend using our data evidence designs as a baseline, tailoring them to the needs of target users and domain characteristics/conventions while adhering to our general design guidance.
For instance, in contexts where ``cherry-picking data''~\cite{Linsnic_CHI23_MisleadingBeyondVisualTricks, walenz2014finding} is a concern, it is recommended to add visual elements that present more contextual information; for example, supplementing a line graph that depicts the \emph{trend} with an \emph{overview} (example shown in \autoref{fig:contextvis}) can provide added context to ``cherry-picking timeframes'', potentially assisting audiences in better judgment about the original claim.

\section{Limitations}\label{sec:limitations}

\subsubsection*{\textbf{Sole focus on fact-level assessment}}
While our method prioritizes verifying the underlying data facts, it is important to acknowledge the potential logical fallacies or misinterpretations that could undermine the plausibility of the resulting conclusions. Consider extending the example presented in \autoref{fig:contextvis} to ``\textit{the unemployment rate experienced a decrease between April 2020 and March 2023, indicating a strong job market}.'' Although the initial part of this data fact may be technically accurate, the concluding inference might not hold true. The claim omits the `spike' in March 2020, coinciding with the COVID-19 pandemic lockdowns in many U.S. states.

\subsubsection*{\textbf{Require manual selection of reference dataset}}
Another limitation of \sys{} is the need for users to manually select or upload suitable reference data for fact-checking. In real-world scenarios, however, data articles or claims may derive from various data sources. Furthermore, the users may have limited access to, or knowledge of, such datasets, which restricts their ability to select appropriate reference data. Although \sys{} currently incorporates features that facilitate this process (i.e., allowing for incorporating additional reference dataset and leveraging GPT to assess the suitability of a dataset for fact-checking), there is room for further enhancement through the adoption of more sophisticated tools. This includes developing more reliable mechanisms to evaluate the suitability of available datasets and automated methods to extract reference data from extensive data infrastructures and reconcile synergies and conflicts among multiple datasets.

\subsubsection*{\textbf{Lack of comprehensive optimization and evaluation of the LLM-based pipeline}} 
We acknowledge that there is potential for further optimization of our NLP pipeline (\autoref{fig:prompt_pipeline}) to enhance its effectiveness and undergo a more comprehensive evaluation. It's noteworthy that the modular structure of our chained LLM-based pipeline offers significant flexibility in the choice of models for downstream tasks. Future work could explore substituting or comparing the LLM-based approach with more established NLP methods for specific steps (e.g., statistical models for coreferences and ellipses resolution~\cite{aralikatte-etal-2021-ellipsis, joshi-etal-2020-spanbert}) and experimenting with different sequences of steps.

\subsubsection*{\textbf{User study designs}}
One study limitation pertains to the demographics of our study participants. While our participants exhibit diverse interactions with and trust in data articles/reports, they are uniformly graduate students who are likely more proficient with data compared to other potential user groups (e.g., fact-check professionals and general news readers). Therefore, we consider our quantitative results to be more relevant to `data-savvy' individuals. Nevertheless, we acknowledge the need for future experiments with other user groups to further broaden the scope of our findings.

\subsubsection*{\textbf{Limitation and risk in adopting LLM}}
The limitations and inherent risks associated with LLMs (e.g., hallucination, inconsistent accuracy) can impact automated fact-checking systems that rely on them. This is particularly notable for tasks like information search~\cite{Shah_situatedsearch} and veracity prediction, especially when contextual information is scarce~\cite{quelle2023perils}. Our approach, in contrast, does not rely on LLM for direct determination of claims' veracity. Instead, the verification is through computing the pertinent data retrieved by data fact specification. While inaccuracies may occur at other stages of the pipeline, such as retrieving an incorrect subset of data or misinterpreting focused attributes, \sys{} offers features that assist in identifying such errors and manually correcting them.

\section{Future Research}\label{sec:futureresearch}

\subsubsection*{\textbf{Assessing and Communicating Plausibility of Data Claims On Reasoning Level}}
Our first limitation highlights the need for developing automated/semi-automated technology that assesses and communicates the plausibility of data claims at both the \emph{factual level} and the \emph{reasoning level}. To achieve this, similarly, a series of NLP tasks are required to determine the linguistic relationship between \emph{data facts} and corresponding \emph{conclusions}, as well as retrieve the data evidence. A reasoning error taxonomy, similar to the visual error typology~\cite{Linsnic_CHI23_MisleadingBeyondVisualTricks} recently proposed by \citeauthor{Linsnic_CHI23_MisleadingBeyondVisualTricks}, is essential for determining the text-data reasoning error and subsequent evidence communication. Given the intricacy of reasoning errors compared to factual ones, we anticipate that visualization --- with its capability to provide context and communicate uncertainty~\cite{Hullman_2020_uncertainty} --- will assume a larger role. Future HCI/Visualization research can delve into and broaden the design space of data evidence presentation, addressing not just factual errors but also reasoning flaws. Further, evaluating data statements' reasoning validity often involves external contextual information, including facts/knowledge or additional quantitative datasets. \citeauthor{kim2024datadive} recently developed an LLM-based interactive tool~\cite{kim2024datadive} to retrieve relevant data associated with data claims. Such tools can potentially be integrated with \sys{}, enabling a more comprehensive evaluation of data claims.

\subsubsection*{\textbf{Fact-checking More Complex Data Claim}}
There is substantial room for further enhancing \sys{}'s capability to effectively handle more complex data claims. A data claim can become more complex when it 1) involves more complicated data operations (e.g., \emph{multiple filters}), 2) encompasses nuanced semantics, and 3) is compound.
Our evaluation (\autoref{eval:transformation}) reveals that parsing data filters is a primary cause of failures in text-to-data mapping. This poses a substantial challenge for handling data facts involving different filters, such as \emph{proportion} type, where not only must all filters be identified but also be categorized (i.e., \emph{focus} or \emph{subspace}) correctly. We hypothesize that a constructive improvement might involve the introduction of a specific step dedicated to subspace information extraction.
Nuanced semantics encompasses a broader range of language descriptors. For example, while \sys{} currently supports two basic types of trends (i.e., increase/decrease), the real-world descriptors for trends can be more diverse and nuanced, considering various adjective/verb pairings (e.g., peak, tanking, spike, etc.)~\cite{bromley2023difference}. Recent studies~\cite{bromley2023difference, Kim_2023_emphasisChecker, Kim_2021_line_chart_caption} on automatic labeling and detecting such visual-text interplay and mismatch regarding temporal data offer more linguistic flexibility and sophisticated computational approaches to quantify such semantic nuances. These studies and approaches could be incorporated into \sys{}, potentially enhancing its capability to verify a wider range of data claims.
As for compound claims, our unoptimized decomposition step can effectively separate simple compound claims. However, when faced with compound data claims coupled with information omission or co-references, \sys{} struggles to decompose them into distinct and accurate data facts, resulting in incomplete data claims or occasional hallucinations. We encourage future research to explore and optimize an end-to-end solution or curate ground-truth datasets for potential task-specific fine-tuning and evaluation.

\section{Conclusion}
Under the backdrop of escalating challenges posed by misinformation, our research delves into data claims --- textual descriptions of facts/insight derived from structured, quantitative data sources. Specifically, we concentrate on two critical problems of automated fact-checking: (1)~retrieving pertinent data evidence to \emph{verify} data claims and (2)~designing effective presentations to \emph{communicate} the data evidence. We developed a prototype, \sys, to operationalize our proposed framework and tackle the multi-faceted challenge. We utilize a pretrained LLM to address the NLP tasks that decompose a data article and transform the data claims into data fact specifications. We explored the design space of data evidence by designing and implementing two representation formats, data table and visualization, across 13 types of data facts. Additionally, we equipped \sys{} with various interactions to enhance its utility and demonstrate its practical potential. Through a performance analysis with a manually curated dataset, we showcased LLM's robust capability both in classifying data facts and in translating textual claims into data fact specifications. We subsequently conducted a mixed-method user study with 20 participants, utilizing our designs as probes to gather insights into \emph{assessment time}, \emph{confidence}, and \emph{preference}. Our findings revealed that our visualization designs are advantageous for 7 out of 13 data fact types regarding \emph{assessment time}. Furthermore, based on participants' feedback and our reflection on the design process, we provided four general design recommendations for presenting data evidence. Ultimately, we discuss the limitations of our study and suggest avenues for future work to adapt and extend our work to accommodate more intricate real-world scenarios and thereby benefit broader audiences.


\balance
\bibliographystyle{ACM-Reference-Format}
\bibliography{bibliography}



\clearpage

\appendix


\section{Appendix}\label{appendix:evidencepresentation}

\begin{figure*}[b]
    \centering
    \includegraphics[width=0.9\textwidth]{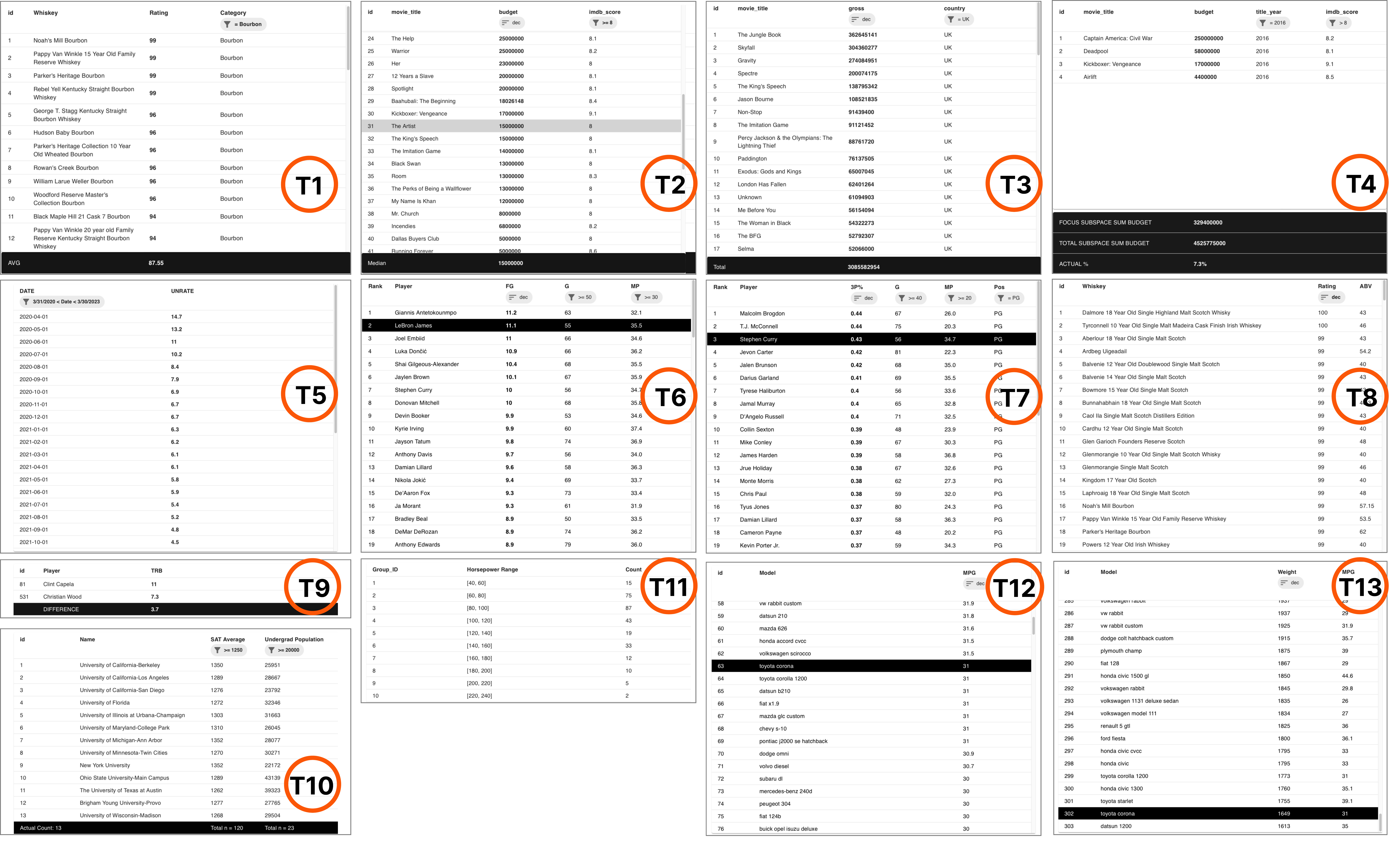}
    \caption{The design of our thirteen data evidence tables. \emph{sorting} and \emph{filtering} widgets are displayed underneath corresponding column names. (T1)~a sticky, highlighted row displaying \emph{mean value}; (T2)~a sticky, highlighted row displaying \emph{median value} and highlighted row(s) for median rows; (T3)~a sticky, highlighted row displaying the \emph{sum (value)}; (T4)~three sticky, highlighted summarization rows displaying sum values of the focus set, the reference set, and the computed \emph{proportion}; (T5)~chronological table showing the mentioned timeframe; (T6)~\emph{extreme} \& (T7)~\emph{rank} - sorted, indexed table highlighting mentioned entity row; (T8)~\emph{association} - two mentioned measures with one measure sorted; (T9)~three-row table displaying the two compared entities and a highlighted row displaying the \emph{difference}; (T10)~sticky, highlighted row displaying the counts of data points satisfying each category and their \emph{categorization} overlap; (T11)~two column table (bins \& range) showing the \emph{distribution}; (T12)~\emph{univariate outlier} \& (T13)~\emph{multivariate outlier} - sorted and indexed table displaying individual values and highlighting mentioned entity row. 
    }
    \Description{}
    \label{fig:table}
\end{figure*}

\begin{figure*}[ht]
    \centering
    \includegraphics[width=0.9\textwidth]{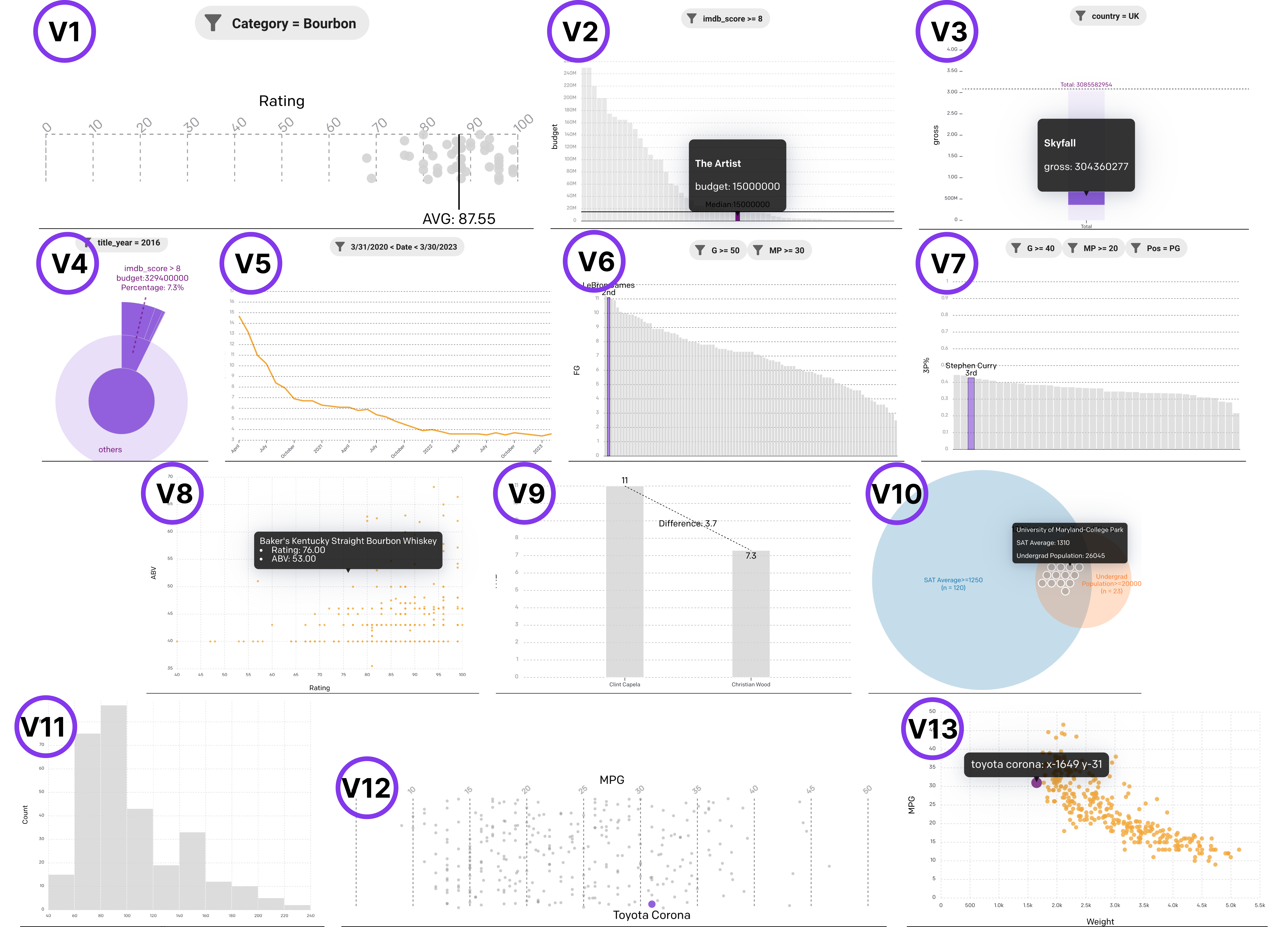}
    \caption{The design of our 13 data evidence visualizations: (V1)~a strip plot accompanied by a line that indicates their \emph{mean value}; (V2)~a sorted bar chart with the \emph{median values} highlighted and labeled; (V3)~a stacked bar chart depicting the \emph{sum (value)}; (V4)~a sunburst plot showing the \emph{proportion} of individual data points; (V5)~a line graph showing the \emph{trend} for a given timeframe; (V6)~\emph{extreme} and (V7)~\emph{rank}: a sorted bar chart with the mentioned data point highlighted and labeled; (V8)~a scatterplot that shows the \emph{association} between values; (V9)~two bars with a comparison line showing the \emph{difference}; (V10)~a proportional Venn diagram showing the number of points based on the \emph{categorization} overlap; (V11)~a histogram displaying the \emph{distribution}; (V12)~\emph{univariate outlier} and (V13)~\emph{multivariate outliers} - a strip/scatter plot with the mentioned data point highlighted. }
    \Description{}
    \label{fig:visexample}
\end{figure*}

\begin{figure*}[ht]
    \centering
    \includegraphics[width=1\textwidth]{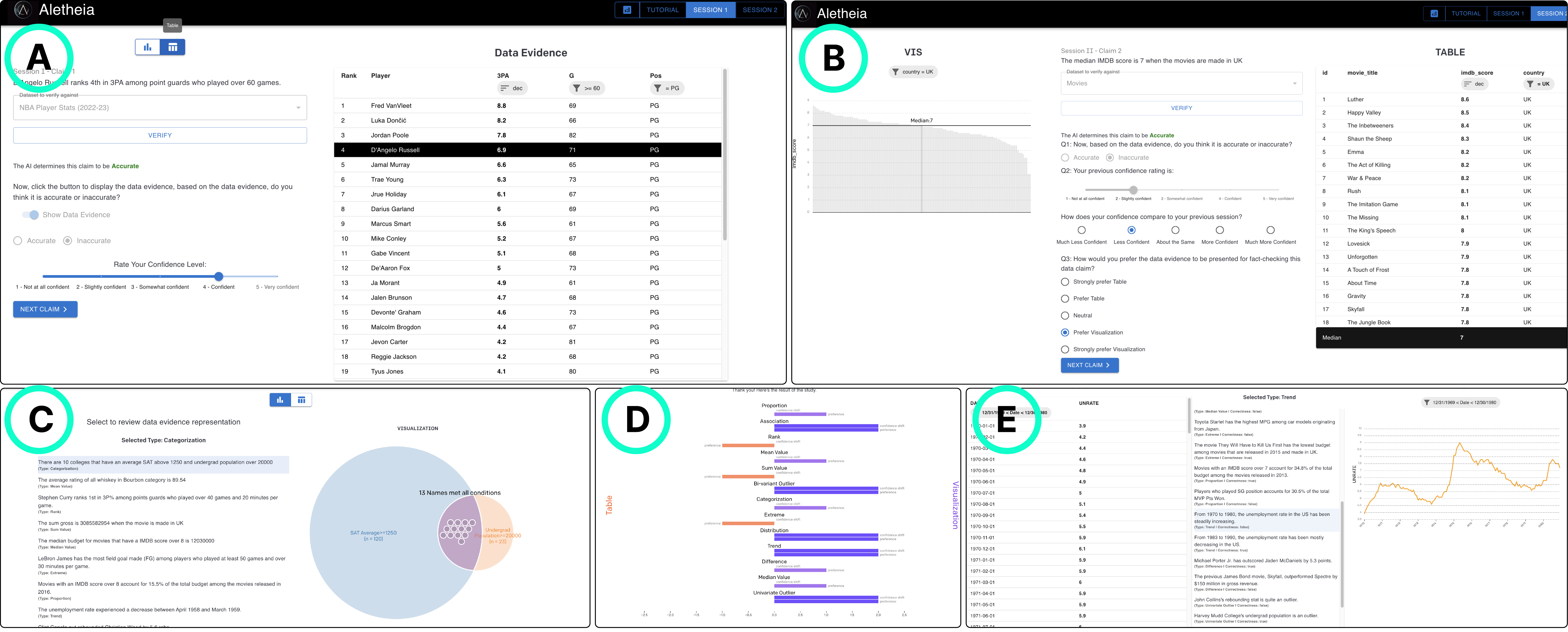}
    \caption{Our user study interface. It consists of five components: (A)~\emph{Phase I page}: the left side includes the questions and selections while the right side displays the respective data evidence representations; (B)~\emph{Phase II page}: its initial state resembles \emph{study I page}. After participants click on \emph{Accurate/Inaccurate} selection, both representations are displayed (left: initial form, right: alternative form). In the middle are the additional questions and scale selections (\ie confidence shift and preference); (C)~\emph{Tutorial page}: it allows participants to click through the demo claims and get familiarized with the encodings for respective representation forms; (D)~\emph{Result chart}: it appears after \emph{Phase II} and demonstrates the results of participants' average confidence shift and preference for each type of data fact; (E)~\emph{Exploration page}: supports participants to click through all the test claims and review the corresponding representations during our interview portion.     
    }
    \Description{}
    \label{fig:userinterface}
\end{figure*}

\end{document}